\documentclass[sigconf,nonacm,balance=false]{acmart}
\usepackage{enumitem}
\usepackage{amsmath}
\usepackage{amsthm}
\usepackage{tikz}
\usepackage[framemethod=TikZ]{mdframed}
\usepackage{caption}
\usepackage{subcaption}
\usepackage{hyperref}
\usepackage{multirow}
\usepackage{algorithm}
\usepackage{algpseudocode}
\usepackage{lipsum}
\usepackage{listings}
\usepackage{xcolor}
\usepackage{tcolorbox}
\tcbuselibrary{breakable, listings, skins}

\usepackage{popets}
\setcopyright{popets}
\copyrightyear{YYYY}

\acmYear{YYYY}
\acmVolume{YYYY}
\acmNumber{X}
\acmDOI{XXXXXXX.XXXXXXX}
\acmISBN{}
\acmConference{Proceedings on Privacy Enhancing Technologies}
\theoremstyle{plain}

\theoremstyle{definition}
\newtheorem{definition}{Definition}[section]

\newtheorem{example}{Example}[section]
\theoremstyle{remark}

\newcommand{\method}{\texttt{ARIEL}}

\newtcblisting[auto counter]{prompt}[2][]{%
  listing only,
  breakable,
  enhanced,
  title={\textbf{Prompt \thetcbcounter:} #2},
  label={#1},                 
  colback=gray!5,
  colframe=gray!50,
  coltitle=black,
  boxrule=0.5pt,
  arc=2pt,
  listing options={
    basicstyle=\ttfamily\small,
    breaklines=true,
    columns=fullflexible,
    mathescape=false
  },
}

\newcommand\todo[1]{\textcolor{purple}{TODO: #1}}

\newif\iftodo
\todotrue
\iftodo
\else
\renewcommand{\todo}[1]{}
\fi

\begin{document}

\title[Personalizing Agent Privacy Decisions via Logical Entailment]{Personalizing Agent Privacy Decisions via Logical Entailment}


\author{James Flemings}
\authornote{Work done while interning at Google Research.}
\affiliation{%
  \institution{University of Southern California}
  \city{}
  \state{California}
  \country{USA}
  }
\email{jamesf17@usc.edu}

\author{Ren Yi}
\affiliation{%
  \institution{Google Research}
  \city{}
  \state{New York}
  \country{USA}
}
\email{ryi@google.com}

\author{Octavian Suciu}
\affiliation{%
  \institution{Google Research}
  \city{}
  \state{New York}
  \country{USA}
  }
\email{osuciu@google.com}

\author{Kassem Fawaz}
\authornote{Work done as a visiting faculty researcher with Google.}
\affiliation{%
 \institution{University of Wisconsin - Madison}
 \city{}
 \state{Wisconsin}
 \country{USA}
 }
\email{kfawaz@wisc.edu}

\author{Murali Annavaram}
\affiliation{%
  \institution{University of Southern California}
  \city{}
  \state{California}
  \country{USA}
}
\email{annavara@usc.edu}

\author{Marco Gruteser}
\affiliation{%
  \institution{Google Research}
  \city{}
  \state{New York}
  \country{USA}
  }
\email{gruteser@google.com}


\renewcommand{\shortauthors}{Flemings et al.}

\begin{abstract}
    Personal large language model (LLM) agents increasingly perform tasks that require access to user data, raising concerns about appropriate data disclosure. We show that relying solely on LLMs to make data-sharing decisions is insufficient. Prompting LLMs with general privacy norms fails to capture individual users’ privacy preferences, while providing prior user data-sharing decisions through in-context learning (ICL) leads to unreliable and opaque reasoning. To address these limitations, we propose \method{} (Agentic Reasoning with Individualized Entailment Logic), a framework that combines LLMs with rule-based logic to enable structured, personalized privacy reasoning. The core mechanism of \method{} determines whether a user’s prior decision on a data-sharing request \textit{logically entails} the same decision for a new request. Experimental evaluations using advanced models and public datasets show that \method{} reduces the F1 error rate for appropriate judgments by \textbf{40.6\%} compared to standard ICL-based reasoning, indicating that \method{} is effective at correctly judging requests where the user would approve data sharing. These results demonstrate that integrating LLMs with logical entailment provides an effective and interpretable approach for automating personalized privacy decisions.
\end{abstract}


\maketitle

\section{Introduction}

\begin{figure}
    \centering
    \includegraphics[width=1.0\linewidth]{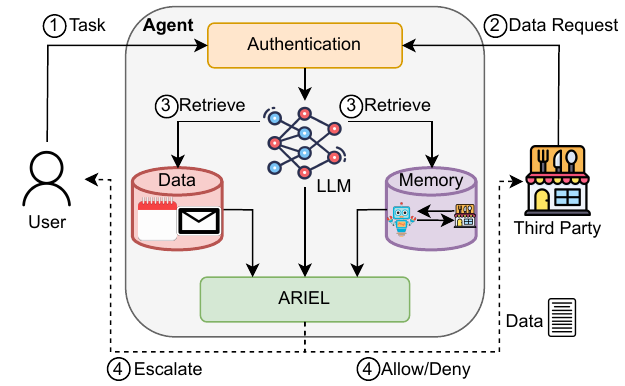}
    \caption{A schematic visualization of the problem setup with \method{}, our proposed privacy-reasoning framework, operating in a larger end-to-end agentic system. \textcircled{1} A user sends a task to the agent to complete. \textcircled{2} A third party requests user data in order to complete the task. \textcircled{3} After the data request has been authenticated, the agent retrieves the relevant user data and prior user judgments on data-sharing requests from memory. \textcircled{4} The agent decides whether it is appropriate to share the user's information to fulfill the third party's request. \method{} either allows or denies transmitting the user data, or escalates the data-sharing request to the user for further assistance.}
    \label{fig:problem_setup}
\end{figure}

As personal AI agents increasingly automate tasks, they frequently encounter scenarios requiring the transmission of private user data. Consequently, the agent must determine the appropriateness of such disclosures. Consider the restaurant booking scenario in Figure \ref{fig:problem_setup}, where a restaurant requests the user's phone number to complete the table reservation. While the agent could seek the user's permission to transmit their phone number, frequent interruptions lead to 'privacy fatigue,' causing disengagement and ineffective data control by the user \cite{acquisti2006there, choi2018role}. To alleviate this cognitive burden, the agent should ideally be capable of autonomously judging when data sharing is appropriate.

Prior to Large Language Models (LLMs), Personal Privacy Assistants (PPAs) relied on either rule-based \cite{zhan2022model, zhan2023privacy} or traditional machine learning (ML) approaches \cite{barbosa2019if, amoros2023predicting}. These systems leverage a history of prior user privacy judgments to decide whether to disclose data for incoming requests. For instance, as shown in Figure \ref{fig:problem_setup}, if a user previously authorized sharing a phone number for a restaurant reservation, the agent can utilize this precedent to approve similar future requests. However, these methods face significant limitations: (1) rule-based approaches are domain-specific and difficult to transfer without substantial modification; and (2) ML methods require extensive user data (from many users) to achieve high performance. 

The integration of LLMs into agentic workflows \cite{wang2024survey} offers a promising solution to these challenges, as their In-Context Learning (ICL) capabilities enable effective generalization even from sparse data~\cite{brown2020language}. While recent studies have evaluated privacy judgments by LLMs~\cite{mireshghallah2024can, shao2024privacylens}, they typically ground their definition of appropriateness in general Contextual Integrity (CI) norms derived from laws or societal consensus \cite{nissenbaum2004privacy, nissenbaum2009privacy}. However, relying solely on general norms fails to capture the nuances of user privacy decisions \cite{barkhuus2012mismeasurement, guo2025not, tran2025understanding, zhang2025towards}. In the example above, it is generally acceptable to share a phone number for a restaurant reservation, yet a user with a restrictive history of data sharing might explicitly reject this request. General norms cannot account for individual privacy decisions, thus being a barrier in the development of personalized privacy agents. This critical gap motivates the central objective of our work.

\begin{quote}
\emph{\textbf{Objective:} Personalizing LLM-based agents to make data-sharing decisions that maximize alignment with prior user decisions, minimize user intervention, and ensure auditability.}
\end{quote}

To fulfill this objective, we identify three requirements for designing personal LLM agents.

\begin{itemize}
    \item \textit{Alignment.} The agent's privacy decisions must adhere to the user's prior privacy judgments, which is important for effective personalization.
    
    \item \textit{Interpretability.} The agent must provide interpretable and traceable reasoning for every privacy judgment, allowing the user to audit the process and maintain trust in the agent's decision-making.

    \item \textit{User Agency.} The agent must preserve user agency by identifying when prior user privacy judgments are insufficient or ambiguous then escalating such requests to the user rather than forcing a judgment. 
    
\end{itemize}

We observe that current LLM-based approaches fail to simultaneously satisfy these design requirements. While leveraging ICL on prior user privacy judgments improves personalization over general privacy norms, it does not achieve effective alignment with prior user privacy decisions due to limited logical reasoning \cite{lin2025zebralogic} and potential for LLM hallucinations \cite{ji2023survey, kalai2025language, zhang2023siren}. Furthermore, the opaque nature of these models undermines interpretability by preventing verification of the reasoning steps used to reach a privacy judgment. Because LLM-generated reasoning traces are often unfaithful to the actual decision process~\cite{barez2025chain, ye2022unreliability}, users cannot effectively audit the agent, ultimately compromising their agency.

Thus, we propose \method{} (Agentic Reasoning with Individualized Entailment Logic), a framework that satisfies these three requirements. Our key insight is to frame personalized privacy judgments as \emph{logical entailment}: determining whether a user's judgment on a prior request entails the same judgment for an incoming request. Rather than relying on LLMs to provide an opaque appropriateness judgment, \method{} combines LLM-based reasoning and rule-based components for reliable and aligned judgments. Specifically, \method{} leverages an LLM to generate ontologies that capture semantic relationships from the user's prior privacy judgments (e.g., a data sensitivity hierarchy), which are then processed by a rule-based component to provide a final judgment. This framework ensures that the agent acts only when a decision can be inferred from the user's prior privacy judgments; otherwise, the request is escalated to the user (Figure \ref{fig:problem_setup}).

We evaluate \method{} against LLM-based reasoning baselines that rely on either general privacy norms or prior user privacy judgments. These evaluations were conducted using state-of-the-art LLMs---including GPT-5, Gemini 2.5 Pro, and Claude Sonnet 4.5---on publicly available datasets covering smart home assistant~\cite{abdi2021privacy} and educational~\cite{shvartzshnaider2016learning} data-sharing contexts. We find that (1) grounding privacy judgments in prior user privacy judgments yields significantly better personalization than relying solely on general privacy norms, and (2) \method{} outperforms pure LLM-based reasoning by reducing the F1 score error of appropriate judgments by up to \textbf{40.6\%}. Our findings indicate that \method{} is effective at correctly judging requests where the user would approve data sharing. We summarize \textbf{our contributions} below:

\begin{enumerate}
    \item We demonstrate the limitations of aligning LLM agents with general privacy norms, as well as the shortcomings of current LLM-based approaches that use prior user judgments for personalized data-sharing decisions.
    
    \item We propose \method{}, a framework for grounding the agent's privacy judgments in prior user privacy judgments through logical entailment with LLM-generated ontologies. Jointly leveraging LLMs and logical entailment overcomes the rigidity of traditional rule-based systems without incurring the inconsistency inherent in pure LLM-based reasoning.
    
    \item We show that \method{} outperforms LLM-based reasoning for personalized privacy decisions, while incorporating user agency and interpretability. Notably, we find that performance gains are larger with smaller models (Gemma 3 4B) and that \method{} remains effective with as few as 20 prior requests.
\end{enumerate}
\section{Related Work}
We review closely-related work on personalizing LLM-based agents to make privacy decisions aligned with user privacy preferences. We discuss (1) prior work in developing personal privacy assistants and (2) evaluations of privacy norms and user preferences in LLMs. Finally, we examine works in (3) neurosymbolic reasoning and their relation to \method{}.

\subsection{Personal Privacy Assistants}
Personal Privacy Assistants (PPAs) aim to reduce the user mental burden in making privacy decisions~\cite{colnago2020informing, stover2023investigating}. They explore automating mobile app permissions, predicting user privacy preferences, and making information-sharing decisions \cite{morel2025ai}. Earlier PPAs targeting mobile app permissions clustered similar user privacy preferences to derive privacy profiles that help determine mobile app permissions for individual users \cite{lin2014modeling, liu2016follow}. Follow-up works utilized contextual factors, such as device type, requesting app, and the permission type, as features for predicting privacy preferences using traditional ML techniques like logistic regression or collaborative filtering \cite{naeini2017privacy, barbosa2019if, amoros2023predicting, wijesekera2017feasibility, wijesekera2018contextualizing}. Other works have proposed PPAs that manage data-sharing decisions by decomposing scenarios into Contextual Integrity (CI) information flows. These systems then utilize rule-based \cite{zhan2022model, zhan2023privacy} or ML-based \cite{kokciyan2022taking} approaches to infer the privacy decision. Finally, Natural Language Inference (NLI)-based models have been proposed to detect mismatches between different privacy policies or between policies and user privacy preferences \cite{alshamsan2022detecting, hosseini2021ambiguity, rao2016expecting}. While significant progress has been made for PPAs, they remain limited in that they require aggregating all user data for training, which requires users to upload their privacy preferences and prior decisions to a centralized entity. Moreover, these methods struggle with generalizing to different domains. 

\subsection{Contextual Privacy Awareness in LLMs}
Recent work has evaluated LLMs on privacy-decision scenarios derived from general privacy norms, while also evaluating alignment of LLM privacy decisions with user privacy preferences.

\paragraph{General Privacy Norms} 
Recently, researchers studied whether LLMs can judge the appropriateness of data-sharing scenarios that are based on CI \cite{nissenbaum2004privacy, nissenbaum2009privacy}. In particular, \citet{mireshghallah2024can} and \citet{shao2024privacylens} developed ConfAIde and PrivacyLens, respectively, which are datasets that evaluate whether LLMs can reason about sharing data. \citet{yi2025privacy} studied ambiguity in these datasets to incorporate contextual expansions for improved privacy reasoning. Furthermore, \citet{lan2025contextual} improved LLM evaluation on PrivacyLens through post-training on a synthetic dataset. More recently, \citet{li2025privaci} developed a benchmark targeting legal compliance. And \citet{mireshghallah2025cimemories} evaluates CI compliance of LLMs with persistent memory that was synthetically generated. These datasets encompass privacy norms sourced from either legal statutes or aggregated judgments from crowdsourced studies, both of which capture the general public's privacy expectations on whether certain data sharing is appropriate in a particular context. Other works have operationalized the contextual awareness of LLMs to reduce privacy leakage of user tasks. \citet{ghalebikesabi2024operationalizing} developed strategies for LLM-based assistants to be CI compliant in form filling tasks. \citet{bagdasarian2024airgapagent} implemented a minimization module with an LLM to generate a minimized set of data based on the user's task. \citet{ngong2025protecting} utilized LLMs to reduce private information within a user's prompt. While evaluating LLMs on general privacy norms is important, these norms fail to capture the nuances of individual user preferences \cite{barkhuus2012mismeasurement, guo2025not, tran2025understanding, zhang2025towards}. Consequently, reliance on general norms alone is ineffective for developing agents tasked with making personalized privacy decisions.

\paragraph{User Privacy Preferences}
\citet{groschupp2025can} and \citet{wu2025towards} investigated personalized access control in LLMs, where the LLM must determine whether to grant data access by evaluating an incoming request against a user's access-control preferences (e.g., "I prefer not to share data unless necessary"). However, it might not be practical to assume that users can explicitly specify access-control preferences. Further, delegating the access control decision to LLMs may result in misalignment with the user's access-control preferences, especially with limitations of LLMs such as hallucinations \cite{ji2023survey, kalai2025language, zhang2023siren} and biases \cite{gallegos2024bias}. Another work studied how well LLMs align with human preferences in household human-robot interaction contexts \cite{sullivan2025benchmarking}. Our work, on the other hand, focuses on a different problem setup, where we evaluate LLMs making privacy decisions on behalf of users. The final work evaluates how well judgments from LLMs align with user expectations \cite{cheng2024ci}. However, their evaluation is quite limited as the dataset is synthetically generated with the authors annotating the user-appropriateness labels. 

\subsection{Neurosymbolic Approaches for Reasoning}
Prior works that are methodologically similar to ours broadly fall under neurosymbolic reasoning, specifically where the LLM acts as an interface to user inputs and offloads the reasoning to rule-based/logical approaches. Such works include using an LLM to translate natural language into a formal intermediate language, which is then offloaded to a deterministic solver to perform the logical reasoning \cite{olausson2023linc, pan2023logic, trinh2024solving, ye2023satlm}. While \method{} does not employ a theorem solver, it does use few-shot prompting with an LLM to map data-sharing requests to levels in a corresponding ontology, which are then passed through a rule-base component to determine entailment. 
\section{Personalizing Privacy Decisions}

As illustrated in Figure \ref{fig:problem_setup}, a personal agent executes tasks on a user's behalf. The agent has access to the user's data and manages data-sharing requests from third parties required to complete these tasks. Below, we formalize the problem setup describing how the agent determines whether transmitting user data is appropriate.

\subsection{Problem Statement}
\label{sec:problem_statement}
We consider an LLM-based agent $A_{\theta}$ that leverages persistent memory $D_u$ \cite{chhikara2025mem0, xu2025mem} to make personalized privacy decisions for a user $u$. We assume a non-adversarial setting where the data-sharing requests $R_u$ are either authenticated or otherwise not maliciously designed to manipulate the agent $A_{\theta}$. Upon receiving a request $R_u$ to share data $d_u$, the agent must infer the user's judgment $l_u$: 
\begin{equation*}
A_{\theta}(D_u, R_u) \in \{\text{appropriate, inappropriate} \}.
\end{equation*}
The memory $D_u$ represents a history of the user's past privacy judgments $l^i_u$ on data-sharing requests $R^i_u$:
\begin{equation}D_{u} = \{P^i_u\}_{i=1}^{n}, \quad \text{where } P^i_u = (R^i_u, l^i_u).
\end{equation}

\smallskip
\noindent
To illustrate the problem setup, consider the following scenario:

\begin{example}\label{ex:requests}
Let the incoming request $R_u$ be from a bank asking for the user's partial Social Security Number (SSN) to open a checking account. The memory $D_u$ contains a prior request $R^i_u$ in which the same bank asked for the user's full SSN for the same purpose, which the user judged as appropriate ($l^i_u = \texttt{appropriate}$). The agent must decide whether it can infer the appropriateness of $R_u$ based on this precedent $(R^i_u, l^i_u)$.
\end{example}

We represent each data request as a CI information flow \cite{nissenbaum2004privacy, nissenbaum2009privacy} containing five parameters: (1) \textit{data type}, the data that is being transmitted; (2) \textit{data subject}, the owner of the data; (3) \textit{data sender}, the individual sharing the data; (4) \textit{data recipient}, the individual receiving the data; (5) \textit{transmission principle}, the purpose and/or condition under which the data is shared. We formally define a request below:

\begin{definition}[Request]\label{def:request}
    Define $R_u = \{d_u, u, s, r, t \}$ as a set containing five parameters where $d_u$ is the user's data, $u$ is the data subject, $s$ is the data sender, $r$ is the data recipient, and $t$ is the transmission principle.
\end{definition}

Using Definition \ref{def:request}, we can map the prior and incoming requests from Example \ref{ex:requests} to the corresponding five parameters, shown in Figure \ref{fig:mapped_requests}.

\begin{figure}[h!]
    \centering
    \begin{subfigure}{0.23\textwidth}
        \begin{mdframed}[backgroundcolor=gray!5, linecolor=black, linewidth=1pt, roundcorner=3pt]
           \begin{center}
               \textbf{Prior Request}
           \end{center} 
        \noindent \textbf{data type:} full SSN \\
        \textbf{data subject:} user \\
        \textbf{data sender:} agent \\
        \textbf{data recipient:} bank \\
        \textbf{transmission principle:} open checking account \\
        \textbf{judgment:} appropriate
        \end{mdframed}
    \end{subfigure}
    \begin{subfigure}{0.23\textwidth}
        \begin{mdframed}[backgroundcolor=gray!5, linecolor=black, linewidth=1pt, roundcorner=3pt]
           \begin{center}
               \textbf{Incoming Request}
           \end{center} 
        \noindent \textbf{data type:} partial SSN \\
        \textbf{data subject:} user \\
        \textbf{data sender:} agent \\
        \textbf{data recipient:} bank \\
        \textbf{transmission principle:} open checking account \\
        \textbf{judgment: }
        \end{mdframed}
    \end{subfigure}
     \caption{A example mapping of the prior request with user judgment and the incoming request from Example \ref{ex:requests} to the five parameters.}
     \label{fig:mapped_requests}
\end{figure}

\subsection{Baseline Approaches}\label{sec:llm_approaches}
Using an LLM to address the privacy decision problem defined above requires addressing a key challenge: how to design the prompt for the LLM. Here, we discuss two baseline approaches to prompt the LLM to make privacy judgments, along with their limitations. 

\paragraph{Privacy Norms-based Baseline.}
The first choice for prompting the LLM is to utilize Privacy Norms. In this approach, the prompt contains three parts. The first part is the instruction to the LLM to make privacy decisions, the second part includes the privacy norms, and the third part is the incoming request. These privacy norms, akin to CI-inspired approaches, can include the \textit{general population's} overall acceptability of a particular action \cite{fan2024goldcoin}. For instance, determining whether a user should share a partial SSN with a bank would rely on broad societal acceptability. However, such an approach is limited because general norms lack the granularity required to capture user-specific, personalized privacy decisions. 

\paragraph{Privacy Preferences-based Baseline.}
To address this limitation, an alternative approach is to prompt the LLM explicitly with the user's privacy preferences. In this approach, the prompt also includes three parts: the general task instruction, the user's privacy preferences, and the incoming request. These preferences can be based on rules or policies that the user specifies, similar to \citet{groschupp2025can}, or they can include a list of the user's privacy judgments on prior requests (e.g., leveraging a past decision about sharing a full SSN to infer the judgment of incoming request) \cite{wijesekera2017feasibility, colnago2022concern}. 

\paragraph{Limitations}
The prompting approaches discussed above, whether based on general norms or prior user privacy decisions, exhibit limitations in terms of (1) alignment, (2) interpretability, and (3) user agency.

Appropriateness judgments generated directly by LLMs remain unreliable due to limited logical reasoning capabilities \cite{lin2025zebralogic} and the potential for hallucinations that deviate from system prompts \cite{ji2023survey, kalai2025language, zhang2023siren}. Consequently, these judgments are not guaranteed to \textit{align} with user privacy preferences, resulting in errors that limit effective personalization.

Furthermore, the decision-making process is opaque. Even when reasoning traces are elicited, they are often unfaithful to the LLM's actual logic \cite{barez2025chain, ye2022unreliability}. This limitation prevents users from accurately \textit{interpreting} the reasoning behind the agent's judgment, thereby eroding the user's trust in the agent.

Finally, because these methods do not align with user preferences and are not inherently interpretable, they hinder \textit{user agency}. While personalized agents should aim to mitigate privacy fatigue, this objective must not come at the cost of compromising user awareness of and control over how decisions are made on their behalf \cite{mitchell2025fully, zhang2025autonomy}.

\subsection{Requirements}\label{sec:requirements}

There is a need to develop agents that make privacy decisions while avoiding the limitations mentioned above. In particular, we posit that such agents must satisfy the following requirements:

\begin{enumerate}
    \item \textbf{Alignment.} The privacy decisions made by the agent on behalf of the user need to accurately reflect the user's prior privacy decisions. Alignment is crucial for effective personalization of agents.
    
    \item \textbf{Interpretability.} The agent needs to provide reasoning that is transparent and traceable so that the user stays in the loop throughout the privacy-decision process. Interpretability maintains user trust in the system.
    
    \item \textbf{User Agency.} In scenarios where the agent lacks enough information from the user's prior privacy decisions to make a judgment, the agent must determine when it should escalate the request to the user for assistance. Importantly, the agent needs to strike the right trade-off between automated privacy decisions and user control.
\end{enumerate}

\section{\method{}: Privacy Judgments with Entailment}\label{sec:ariel}

We propose \method{}, a framework enabling aligned, structured, and interpretable reasoning for personalized privacy judgments. \method{} operates as a specialized module for privacy decisions within an end-to-end agentic system (see Figure~\ref{fig:problem_setup}). We describe the framework in detail below.

\begin{figure*}[t!]
    \centering
    \includegraphics[width=0.8\linewidth]{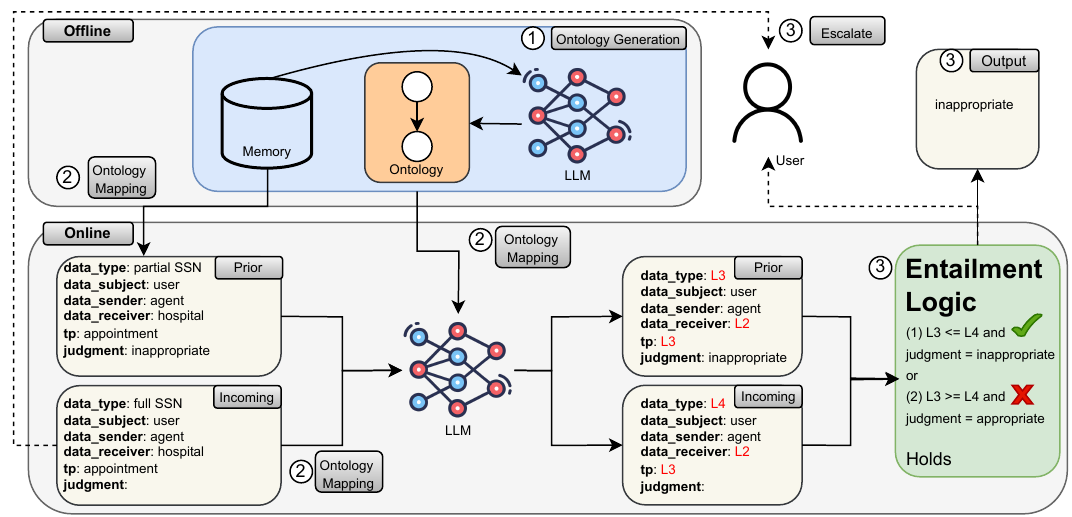}
    \caption{High-level overview of \method{}, which can be broken down into three components. \textcircled{1} \method{} uses an LLM to generate an ontology for each user based on the user's prior privacy judgments on data-sharing requests. \textcircled{2} Once \method{} receives an incoming request, it goes through each prior request from memory to determine if entailment holds. The LLM maps the parameters in both requests to corresponding levels in the generated ontologies. \textcircled{3} A set of rules is applied to determine whether entailment holds between each mapped prior request and the mapped incoming request. If the incoming request is not entailed by any prior requests, \method{} escalates the incoming request to the user.}
    \label{fig:proprosed_framework}
\end{figure*}

\subsection{Overview}
\label{sec:overview}

\subsubsection{Entailment}

The core principle of \method{} is to personalize privacy judgments by determining if a prior judgment made by the user necessitates the same judgment for the incoming request:
\begin{center}
$(R^i_u, l^i_u) \implies (R_u, l^i_u)$ \textit{Does the judgment $l^i_u$ for the prior request $R^i_u$ logically imply the same judgment for the incoming request $R_u$?}
\end{center}

This approach addresses the limitations of LLM-based baselines, which do not provide a logical way to infer user judgments from prior decisions. In contrast, entailment treats prior judgments as logical premises that constrain future decisions, offering an interpretable and controllable method for making privacy decisions. The following example highlights how entailment enables privacy decisions based on prior requests. 

\begin{example}\label{ex:entailment}
    Referring back to Example \ref{ex:requests}: a user previously approved sharing their \textit{full} SSN to open a bank account ($l^i_u = \texttt{appropriate}$). The incoming request asks for a \textit{partial} SSN in the exact same context. Logically, the user's approval to share the more sensitive data (full SSN) entails approval to share the less sensitive data (partial SSN).
\end{example}

The main challenge toward applying entailment is the semantic gap between incoming and prior requests, especially if there is no exact lexical overlap between the parameters of the prior and incoming requests. \method{} addresses this gap using LLM reasoning and rule-based components unified by ontologies. Drawing on prior work in mobile privacy analysis \cite{andow2019policylint, andow2020actions, wang2018guileak}, we utilize ontologies to capture entity relationships between parameters in the incoming and prior requests. For example, we can capture the hierarchy of data sensitivity within an ontology where partial SSN is less sensitive than full SSN. However, we simplify the standard graph structure of ontologies into directed paths, where each vertex represents an entity level (detailed in Section \ref{sec:ontology_gen}).

\subsubsection{Design} The architecture of \method{}, illustrated in Figure \ref{fig:proprosed_framework}, consists of three components: 
\begin{enumerate} 
    \item \textbf{Ontology Generation:} use an LLM to generate relevant ontologies for each user over the parameters in each prior request and incoming request. 

    \item \textbf{Ontology Mapping:} map the parameters of the incoming request and relevant prior requests to levels in the generated ontologies. 

    \item \textbf{Entailment Logic:} apply a set of entailment rules to determine if the user's prior judgments entail the incoming request. 
\end{enumerate}

The three components of \method{} operate in two phases. In the \textbf{offline phase}, \textit{Ontology Generation} creates ontologies for each user based on the user's prior privacy judgments and the LLM's societal norm awareness. The \textbf{online phase} processes each incoming request: first, \textit{Ontology Mapping} selects relevant prior requests with user judgments from memory, then maps the parameters in each request to the generated ontologies. This step provides a unified representation of the parameters in both the incoming and prior requests, enabling the entailment step. Next, \textit{Entailment Logic} applies a set of rules to determine if entailment holds between each mapped prior request and the mapped incoming request. If the incoming request is not entailed by any prior requests, \method{} escalates the incoming request to the user.

\subsubsection{Meeting Requirements} \method{} satisfies the design requirements outlined in Section \ref{sec:requirements} as follows: 

\begin{enumerate} 
    \item \textbf{Alignment:} Appropriateness judgments are grounded strictly in the user's prior privacy judgments via logical entailment. The prior requests are mapped onto ontologies, which are generated from the user's prior privacy judgments. 

    \item \textbf{Interpretability:} All decisions are transparent and auditable. Users can trace \method{}'s judgment back to the specific prior request and the logical rules used to establish the entailment. 

    \item \textbf{User Agency:} \method{} maintains user agency by escalating decisions only when an incoming request is not logically entailed by prior judgments. This prevents the agent from making appropriateness judgments when the user's prior privacy judgments are insufficient. Furthermore, by providing interpretable and aligned privacy judgments, \method{} ensures users retain awareness of and control over the automated decision-making process.
\end{enumerate}

\subsection{Ontology Generation}
\label{sec:ontology_gen}

We first describe how \method{} obtains a set of ontologies $O$. The \textbf{challenge} in generating ontologies is that the agent lacks access to all possible values for each parameter in a request, and it is infeasible to anticipate all these values. For example, a \textit{data recipient} ontology must be sufficiently comprehensive to allow \method{} to accurately map various recipients within a request to the ontology.

\textbf{To address this challenge}, \method{} relies on limited initial information: the domain of the incoming request (e.g., an education context) and the user's prior privacy judgments $D_u$. Leveraging these two sources, the LLM generates user-specific ontologies during an offline phase. The LLM's ability to construct ontologies from limited information enables \method{} to adapt to new domains without substantial modification.

\method{} instructs the LLM to generate an ontology for each request parameter, excluding the \textit{data subject}, which is always the same for each request. The LLM generates each ontology based on a hierarchical relationship. We detail each ontology with its corresponding relationship below (please refer to Appendix \ref{app:method_prompts} for the exact ontology generation prompts):

\begin{enumerate}
    \item \textit{Data Type.} Represents data sensitivity as perceived by the user. Sensitivity \textbf{increases} at higher levels (Level 1 is least sensitive).
    
    \item \textit{Data Sender.} Represents the sender's authority to transmit data. Authority \textbf{decreases} at higher levels (Level 1 has most authority).
    
    \item \textit{Data Recipient.} Represents the user's trust in and the authority of the recipient. Trust and authority \textbf{decrease} at higher levels (Level 1 has highest trust/authority).
    
    \item \textit{Transmission Principle.} Represents the transmission purpose and safeguards. Safeguard strength and purpose relevance \textbf{decrease} at higher levels (Level 1 has the strongest safeguards/relevance).
\end{enumerate}

Each hierarchical relationship is derived from two sources: (1) \textbf{User's prior privacy judgments:} Relationships are inferred directly from the user's prior privacy judgments ($D_u$). For instance, if a user approves a request involving parents but denies an identical request involving friends, \method{} infers a higher trust level for parents. This ensures the ontology is personalized. (2) \textbf{General norms:} For relationships not present in $D_u$, \method{} leverages the LLM's societal norm awareness. For example, the LLM can infer that a typical user generally trusts family more than acquaintances. Figure \ref{fig:example_ontology} illustrates a data type ontology generated by Gemini 2.5 Pro for a particular user in the evaluation dataset using this method.

\begin{figure}[h!]
    \centering
    \begin{subfigure}{0.45\textwidth}
        \begin{mdframed}[backgroundcolor=gray!5, linecolor=black, linewidth=1pt, roundcorner=3pt]
           \begin{center}
               \textbf{\textit{data type} Ontology}
           \end{center} 
        \textbf{L1.} Non-personal, publicly available information. \\
        \textbf{L2.} User preferences and habits that are not directly identifiable. \\
        \textbf{L3.} Information about the user's home environment and security. \\
        \textbf{L4}. Contact information and social connections. \\
        \textbf{L5.} Sensitive personal and location information.  \\
        \textbf{L6.} Highly sensitive health and medical data. \\
        \end{mdframed}
    \end{subfigure}
    \caption{Example of generated \textit{data type} ontology from Gemini 2.5 Pro.}
    \label{fig:example_ontology}
\end{figure}

\subsection{Ontology Mapping}\label{sec:ontology_map}
In the online phase, \method{} evaluates an incoming request $R_u$ by comparing it against prior requests. The \textbf{challenge} is that the privacy contexts can diverge significantly if requests differ by multiple parameters simultaneously, making direct entailment unreliable. For example, sharing full SSN with a bank to open a checking account is an entirely different context than sharing full SSN with an IRS agent to file taxes. Furthermore, even when requests are similar, the agent requires a structured process for comparing differing parameter values based on some hierarchical relationship in order to determine the entailment (e.g., comparing \textit{data recipients} of parents vs friends).

\textbf{To address these challenges}, \method{} employs a structured approach leveraging the generated ontologies $O$, as detailed in Algorithm \ref{alg:ontological_entailment}:

\textbf{Neighboring Requests.} To ensure that comparisons are made between contextually similar requests, we first select prior requests that 'neighbor' the incoming request, denoting the neighboring set as $E$. More precisely, $E$ contains prior requests $R^i_u$ whose Hamming distance $d$ from the incoming request $R_u$ is at most one. This is equivalent to ensuring that both requests differ by only one parameter (line \ref{line:neighboring_requests} in Algorithm \ref{alg:ontological_entailment}). 

\textbf{Ontology Mapping.} To rigorously compare the differing parameters, we introduce an operator $\leq_O$ to formalize the relationship between the parameters of two distinct requests, $R$ and $R'$, based on their mapped levels within $O$. Specifically, $R' \leq_O R$ holds if every parameter in $R'$ is ontologically "less than or equal to" the corresponding parameter in $R$. We formally define this as:
\begin{definition}[Ontological Relationship]\label{def:ontology_relationship}
    Let $R=\{d_u, u, s, r, t\}$, $R' = \{d'_u, u, s', r', t'\}$ be two requests defined in Definition \ref{def:request} that are partially ordered by the set of ontologies $O$. Then $R' \leq_O R$ holds if, for each corresponding parameter $p \in R$ and $p' \in R'$, either $p' <_o p$ or $p' =_o p$ under the relevant ontology $o \in O$.
\end{definition}
Then \method{} iterates through each neighboring prior request $R'_u$. For each pair, the agent retrieves the differing parameter values, $p$ (from $R_u$) and $p'$ (from $R'_u$). These values are then mapped to the relevant ontology $o \in O$ (line \ref{line:ontological_mapping} in Algorithm \ref{alg:ontological_entailment}). For this ontological mapping, we use the LLM $\theta$ to determine the relevant ontology $o$ and map $p$ and $p'$ to their corresponding levels in $o$.

\subsection{Entailment Logic}
Once the requests are mapped to the ontology, \method{} applies a set of rules to determine the judgment of $R_u$. This involves comparing the mapped levels between the incoming request and each prior request using $\leq_O$, and considering the user judgment $l^i_u$ on the prior request $R^i_u$. We formally define our rule-based approach for determining entailment below: 

\begin{definition}[Entailment]\label{def:entailment}
   Let $R, R'$ be two requests with $l'$ being an appropriateness label for $R'$. Let $O$ be a set of ontologies. We say that the entailment $(R', l') \implies (R, l')$ holds if one of the following two cases holds:
   \begin{enumerate}
       \item $l' = \texttt{inappropriate}$ and $R' \leq_O R$ 
       \item $l' = \texttt{appropriate}$ and $R \leq_O R'$
   \end{enumerate}
\end{definition}

\begin{algorithm}[!t]
   \caption{Online phase of \method{}}
   \label{alg:ontological_entailment}
   \begin{algorithmic}[1]
   \Require LLM $\theta$, user $u$, user knowledge base $D_u$, incoming request $R_u$, set of ontologies $O$
   \Ensure \texttt{appropriate}, \texttt{inappropriate}, or \texttt{undetermined} 
        \State $E = \{ \}$ \Comment{\textcolor{blue}{Set of prior request neighboring $R_u$}}
        \State $c_{+} \gets 0, c_{-} \gets 0$
         \For {$(R^i_u, l^i_u)\in D_u$} 
            \If {$d(R^i_u, R_u) \leq 1$} \Comment{\textcolor{blue}{Requests differ by one parameter}} \label{line:neighboring_requests}
                \State $E = E \bigcup \{(R^i_u, l^i_u)\}$
            \EndIf
         \EndFor
        \For {$(R_u', l'_u) \in E$}
            \State Let $p \in R_u$, $p' \in R'_u$ such that $p \neq p'$ \Comment{\textcolor{blue}{Differing parameter}}
            \State $p_O \gets \theta(O, p)$, $p'_O \gets \theta(O, p')$ \label{line:ontological_mapping} \Comment{\textcolor{blue}{Ontological mapping}}
            \If {$l'_u == \texttt{inappropriate} \wedge p'_O \leq p_O$} \label{line:entailment}
                \State $c_{-} \gets c_{-} + 1$ 
            \ElsIf {$l'_u == \texttt{appropriate} \wedge p_O \leq p'_O$ }
                \State $c_{+} \gets c_{+} + 1$
            \EndIf
        \EndFor
        \If {$c_{+} > c_{-}$} \label{line:count_appropriateness} \Comment{\textcolor{blue}{Determine Judgment}}
            \State \textbf{Return} \texttt{appropriate}
        \ElsIf {$c_{-} > c_{+}$} 
            \State \textbf{Return} \texttt{inappropriate}
        \Else  
            \State \textbf{Return} \texttt{Undetermined} \Comment{\textcolor{blue}{No majority, undetermined}}
        \EndIf
   \end{algorithmic} 
\end{algorithm}

\method{} checks if entailment holds using Definition \ref{def:entailment} (line \ref{line:entailment} in Algorithm \ref{alg:ontological_entailment}). We note that multiple neighboring requests can be entailed, with potentially opposing judgments. This could be due to inconsistency in the user's prior judgments or incomplete information from the ontology. To address this, the agent goes through each neighboring request and keeps a running vote of the appropriate and inappropriate judgments that were entailed. Whichever appropriateness label receives the majority vote is the final judgment. If there is no majority, the agent outputs 'undetermined' (line \ref{line:count_appropriateness} in Algorithm \ref{alg:ontological_entailment}), signifying that the incoming request should be escalated to the user for further assistance. 

Lastly, we note that in scenarios where multiple appropriate and inappropriate prior requests are entailed, \method{} can be modified so that the agent to escalate the incoming request to the user instead of employing a majority vote. 
\section{Evaluation Setup}

We present the experimental framework designed to assess the efficacy of \method{}. In the following sections, we detail the specific datasets used, describe the baselines compared, and define the evaluation metrics employed to quantify performance.

\subsection{Datasets for Data-sharing Requests}
\label{sec:datasets}

We use existing datasets that elicit privacy preferences through factorial vignette surveys based on CI. Participants are presented with data-sharing scenarios and asked to indicate their acceptance of each one \cite{abdi2021privacy, apthorpe2018discovering, apthorpe2019evaluating, bhatia2018empirical, hoyle2020privacy, martin2016measuring, shvartzshnaider2016learning}. To simulate a user with prior privacy judgments, each data-sharing scenario is converted into a request (Definition \ref{def:request}) with appropriateness judgments directly received from the participants. Then, we divided a participant's set of answered requests into two pairwise disjoint subsets. The first subset was treated as observed prior requests with the user's answers as the judgments. The second subset represents incoming requests where the participant's answers were held out as the ground truth for evaluation. Our evaluation employs two such datasets, which we repurposed for our setup as detailed below. See Figure \ref{fig:example_requests} for example requests from both datasets. 

\subsubsection{Smart Home Personal Assistant (SPA) dataset} 

The SPA dataset\footnote{Dataset can be accessed here: \url{https://osf.io/63wsm/overview}} \cite{abdi2021privacy} consists of factorial vignette surveys covering different data-sharing scenarios to understand privacy norms in the SPA ecosystem. The scenarios typically involve a voice assistant provider sending user data to individuals/entities who have access to the user's SPA ecosystem, under a certain purpose and condition. The data subject of a request is always "user" and the data sender is always "assistant provider". The total number of participants is 1,730, with each participant randomly assigned to answer a subset of the questions. This resulted in approximately 180 different scenarios answered by each participant.

Due to the relatively large number of participant responses from the SPA dataset, we restrict our evaluation to a set of randomly sampled 500 users. To ensure a consistent experimental setup across all users, we standardized the response size by randomly selecting 70 responses per user. For each of the 500 users, we partitioned the responses into 60 prior requests and 10 incoming requests. This yields a total of $500 * 10 = 5000$ incoming requests, consisting of 1,765 appropriate and 3,235 inappropriate labels. We also perform an ablation study on varying the number of prior requests in Section \ref{sec:ablation}.

Lastly, in the dataset, the participant responses are on a 1-5 Likert scale where 1 and 2 correspond to strongly and somewhat unacceptable responses, respectively, 3 corresponds to neutral, and 4 and 5 correspond to somewhat and strongly acceptable responses, respectively. We convert the responses with 1-2 scores to "Inappropriate" and the responses with 4-5 to "Appropriate". We filter the remaining responses with scores of 3.

\subsubsection{Education Dataset.} The Education dataset\footnote{Dataset can be accessed here: \url{https://yansh.github.io/papers/HCOMP/}} \cite{shvartzshnaider2016learning} consists of factorial vignette surveys involving the transmission of student data within an educational context. For example, a student's grades are shared by a professor to parents if the student is performing poorly. The data subject of a request is always 'student'. In total, there were 451 respondents, each answering a random subset of the questions which was around 88 and 89. 

The participant responses in the dataset contain numerical values, where 1 corresponds to Yes, 2 corresponds to No, and 3-5 correspond to Does not Make Sense options. Specifically, 3 is "The sender is unlikely to have the information", 4 is "The receiver would already have the information", and 5 is "the question is ambiguous." We convert responses with 1 to "Appropriate" and with 2 to "Inappropriate". We filter out the remaining responses with 3-5 scores. Lastly, we filter out users who have fewer than 70 appropriateness judgments, so each user has 60 prior requests and 10 incoming requests. In total, there are 302 users and $302 * 10 = 3020$ total incoming requests, consisting of 1378 appropriate labels and 1642 inappropriate labels.

We present a summary of the statistics of the evaluation datasets after our pre-processing in Table \ref{tbl:dataset_summary}.

\begin{table}[t!]
\resizebox{\linewidth}{!}{%
\begin{tabular}{lp{15mm}p{18mm}lll}
\toprule
\textbf{Dataset}  & \textbf{\# Sampled \newline Users} & \textbf{\# Requests \newline per User} & \textbf{Total} & \textbf{\# App} & \textbf{\# Inapp} \\
\midrule
SPA         & 500      & 10 incoming \newline 60 prior & 5000  & 1765 & 3235             \\
Education & 302      & 10 incoming \newline 60 prior & 3020  & 1378 & 1642 \\
\bottomrule
\end{tabular}}
\caption{A summary of the statistics of the evaluation datasets after our pre-processing. This contains the sampled number of users (\# Users), sampled number of requests per user (\# Requests per User), the total number of incoming requests among all users (Total), the number of incoming requests with appropriate labels (\# App), and the number of incoming requests with inappropriate labels (\# Inapp).}
\label{tbl:dataset_summary}
\end{table}

\begin{figure}[h!]
    \centering
    \begin{subfigure}{0.23\textwidth}
        \begin{mdframed}[backgroundcolor=gray!5, linecolor=black, linewidth=1pt, roundcorner=3pt]
           \begin{center}
               \textbf{SPA Dataset}
           \end{center} 
        \noindent \textbf{data type:}  email content \\
        \textbf{data subject:} user \\
        \textbf{data sender:} assistant provider \\
        \textbf{data recipient:} partner \\
        \textbf{transmission principle:} reading emails on user's behalf \\
        \end{mdframed}
    \end{subfigure}
    \begin{subfigure}{0.23\textwidth}
        \begin{mdframed}[backgroundcolor=gray!5, linecolor=black, linewidth=1pt, roundcorner=3pt]
           \begin{center}
               \textbf{Education Dataset}
           \end{center} 
        \noindent \textbf{data type:}  grades \\
        \textbf{data subject:} student \\
        \textbf{data sender:} professor \\
        \textbf{data recipient:} parents \\
        \textbf{transmission principle:} if student is performing poorly \\ \\
        \end{mdframed}
    \end{subfigure}
    \caption{Example requests from the SPA and Education Dataset that are provided to the LLMs in our evaluations.}
    \label{fig:example_requests}
\end{figure}

\subsection{Baselines}\label{sec:baselines}
Following the description in Section \ref{sec:llm_approaches}, we employ two natural baseline prompt strategies: (1) privacy norms representative of the general population and (2) prior privacy judgments from users. 

\subsubsection{Privacy Norms-based Baseline}\label{sec:prompts_priv_norms}
Privacy norms capture the general population's accepted appropriateness of information flows. We consider two prompts. 

\begin{enumerate}[leftmargin=*]
    \item \textit{zero-shot}. We construct the prompt to utilize the out-of-the-box privacy norm awareness of LLMs, following from recent prior work on evaluating CI-compliance in LLMs \cite{mireshghallah2024can, shao2024privacylens}. The prompt directs the LLM to judge the appropriateness of each information flow based on its knowledge of general privacy norms. The prompts can be found in Appendix \ref{app:zero_shot_prompts}.
    
    \item \textit{privacy norms}. We construct the prompt such that the LLM's privacy judgment is now based on a provided subset of representative privacy norms. These extracted privacy norms are taken directly from the analysis provided in the original papers for each dataset. Table \ref{tbl:general_norms} contains a subset of the general privacy norms from both evaluation datasets that were used by the LLMs. The prompt directs the LLM to judge the appropriateness of each information flow based on the provided general privacy norms. The prompts containing the entire set of privacy norms can be found in Appendix \ref{app:priv_norms_prompts}.
\end{enumerate}

\begin{table}[t!]
\begin{tabular}{p{5cm}l}
\toprule
\textbf{Rule} & \textbf{Judgment} \\ \hline
     \multicolumn{2}{c}{\textbf{SPA Dataset, user recipients}}            \\ \hline
\textbf{data type}: voice command history & Inappropriate \\
\textbf{data type}: bank account details & Inappropriate \\
\textbf{recipient}: neighbors & Inappropriate \\
\textbf{recipient}: visitors in general & inappropriate  \\ \hline
     \multicolumn{2}{c}{\textbf{SPA Dataset, non-user recipients}}            \\ \hline
\textbf{data type}: bank account details     & Inappropriate \\
\textbf{recipient}: advertising agency     & Inappropriate \\
\textbf{tp}: no purpose/condition & Inappropriate  \\ \hline
     \multicolumn{2}{c}{\textbf{Education Dataset}}            \\ \hline
\textbf{sender}: professor; \textbf{recipient}: graduate schools; \textbf{data type}: grades; \textbf{tp}: with subject's consent & Appropriate \\
\textbf{sender}: TA; \textbf{recipient}: classmates; \textbf{data type}: grades; \textbf{tp}: subject performing poorly & Inappropriate
     \\ \bottomrule
\end{tabular}
\caption{A subset of the top privacy norms for user and non-user recipients from the SPA dataset, and the top appropriate and inappropriate norms from the Education dataset. The LLMs are provided either the general norms from the SPA or education dataset, depending on which dataset is being evaluated. We denote tp as transmission principle.}
\label{tbl:general_norms}
\end{table}

\subsubsection{Privacy Preferences-based Baseline}\label{sec:personalization}
We provide privacy judgments on prior requests made by a user. We consider two prompts. 

\begin{enumerate}[leftmargin=*]
    \item \textit{In-Context Learning (ICL)}. We construct the prompt to leverage ICL \cite{brown2020language} by instructing the LLM to infer the user's privacy preferences from their judgments on prior requests $D_u$. When the LLM judges an incoming request $R_u$, its decision is therefore personalized based on the inferred preferences in $D_u$. These prompts are in Appendix \ref{app:icl_prompts}.
    
    \item \textit{ICL (w Undet)}. We modify the instruction in \textit{ICL} to output 'undetermined' if the model is unable to confidently judge whether the incoming request is appropriate or inappropriate based on the user's privacy preferences. This modification accounts for scenarios where the user's privacy preferences may not always provide sufficient information to accurately judge the incoming request. The prompts are in \ref{app:icl_w_undet_prompts}.
\end{enumerate}

\subsection{Models and Evaluation metrics}\label{sec:models_eval_metrics}
\paragraph{Models} We evaluate \method{} and the baselines on three advanced models: Gemini 2.5-Pro (gemini-2.5-pro), GPT-5 (gpt-5-2025-08-07), and Claude Sonnet 4.5 (claude-sonnet-4-5-20250929). Additionally, we include a smaller, open-source model, Gemma 3 4B IT \cite{team2025gemma}.

\paragraph{Metrics} We compare the agent's predicted judgments against each sampled user's judgments on incoming requests. We report separate F1 scores for both the \textit{appropriate} and \textit{inappropriate} classes. Given an evaluation dataset $D$ containing incoming requests $R_u$ with judgments $l_u$ for each user $u$, we define the precision $P_{A}$ for the appropriate class ($c^+ = \texttt{appropriate}$) as:
\begin{equation*}
    P_{A} := \frac{|(R_u, l_u)\in D : A_{\theta}(D_u, R_u) = c^+ \wedge l_u = c^+|}{|(R_u, l_u)\in D : A_{\theta}(D_u, R_u) = c^+|}.
\end{equation*}
We also define recall $R_{A}$ for the appropriate class as:
\begin{equation*}
    R_{A} := \frac{|(R_u, l_u)\in D : A_{\theta}(D_u, R_u) = c^+ \wedge l_u = c^+|}{|(R_u, l_u)\in D : l_u = c^+|}.
\end{equation*}
Then we can define the F1 score $F1_{A}$ for the appropriate class as:
\begin{equation*}
    F1_{A} = 2 \cdot \frac{P_{A} \cdot R_{A}}{P_A + R_A}.
\end{equation*}
A similar process can be used to obtain the F1 score for the inappropriate class $F1_{I}$. High $F1_{A}$ indicates the agent is correctly sharing user data (utility), whereas high $F1_{I}$ indicates the agent is correctly withholding data (privacy). 

\textit{Evaluation Procedure for general norms.} For each unique request in the datasets, we generate a prompt as described in Section \ref{sec:prompts_priv_norms}. This results in 825 prompts for the SPA dataset and 1411 for the Education dataset, covering both \textit{zero-shot} and \textit{privacy norms}. Once the prompts are generated, we provide all of them to the models to obtain their responses. Then, for each user, we compare their judgment on an incoming request with the LLM's judgment on that same request across both general norms prompt types. 

\textit{Evaluation Procedure for user privacy preferences.} We generate a prompt for each incoming request with the user's privacy judgments on prior requests. This results in 5000 prompts for the SPA dataset and 3020 prompts for the Education dataset, covering \textit{ICL}, \textit{ICL (w Undet)}, and \method{}. Incoming requests resulting in an 'undetermined' judgment are excluded from the evaluation for the respective baselines (i.e., \textit{ICL (w Undet)} or \method{}). To account for this partial coverage, we report the Support (Sup) metric, defined as the total number of appropriateness judgments (i.e., "appropriate" or "inappropriate") made by the model. 
\section{Results}\label{sec:user_pref_eval}

\begin{table*}[t!]
\centering
\resizebox{\linewidth}{!}{%
\begin{tabular}{l|c cc cc ccc ccc ccc ccc}
\toprule
\multirow{2}{*}{\textbf{Dataset}} & \multirow{2}{*}{\textbf{Model}} & \multicolumn{2}{c}{\textbf{zero-shot}} & \multicolumn{2}{c}{\textbf{privacy norms}} & \multicolumn{3}{c}{\textbf{ICL}} & \multicolumn{3}{c}{\textbf{ICL (w Undet)}} & \multicolumn{3}{c}{\textbf{\method{}}}
     \\
    \cmidrule(lr){3-4} \cmidrule(lr){5-6} \cmidrule(lr){7-9} \cmidrule(lr){10-12} \cmidrule(lr){13-15} 
    & & $F1_{\text{A}}$ (\%) & $F1_{\text{I}}$ (\%) & $F1_{\text{A}}$ (\%) & $F1_{\text{I}}$ (\%) & $F1_{\text{A}}$ (\%) & $F1_{\text{I}}$ (\%) & Sup & $F1_{\text{A}}$ (\%) & $F1_{\text{I}}$ (\%) & Sup & $F1_{\text{A}}$ (\%) & $F1_{\text{I}}$ (\%) & Sup \\ 
    \midrule
    \multirow{4}{*}{SPA} & Gemini 2.5 Pro & 49.4  & 75.8 & 59.1 & 71.5 & 81.8 & 91.0 & 5000 & 82.4 & 92.2 & 4613 & \textbf{87.2} & \textbf{94.2} & 3797 \\
    & GPT-5 & 27.1 & 78.3 & 58.4 & 71.1 & 79.0 & 90.6 & 5000 & 80.0 & 91.1 & 4921 & \textbf{87.4} & \textbf{94.2} & 3810 \\
    & Claude Sonnet 4.5 & 40.0 & 75.0 & 57.7 & 73.6 & 81.4 & 91.0 & 5000 & 83.3 & 91.7 & 4900 & \textbf{86.6} & \textbf{93.8} & 3824 \\
    & Gemma 3 4B IT & 45.1 & 72.8 & 46.4 & 73.1 & 64.1 & 84.4 & 5000 & 62.5 & 84.2 & 4999 & \textbf{75.5} & \textbf{88.4} & 3680 \\
    \midrule
    \multirow{4}{*}{Education} & Gemini 2.5 Pro & 60.1 & 61.7 & 38.5 & 71.6 & 69.9 & 80.8 & 3020 & 75.5 & 87.2 & 2325 & \textbf{84.1} & \textbf{88.2} & 1681 \\
     & GPT-5 & 62.9 & 55.6 & 51.2 & 69.2 & 74.0 & 82.9 & 3020 & 74.4 & 83.2 & 2961 & \textbf{84.8} & \textbf{89.2} & 1603 \\
     & Claude Sonnet 4.5 & 61.4 & 63.0 & 62.3 & 62.5 & 75.4 & 82.8 & 3020 & 75.0 & 82.8 & 2950 & \textbf{84.0} & \textbf{88.4} & 1608 \\
     & Gemma 3 4B IT & 64.1 & 18.0 & 46.7 & 66.8 & 61.7 & 78.7 & 3020 & 63.4 & 78.8 & 3020 & \textbf{80.4} & \textbf{86.1} & 1644 \\
\bottomrule
\end{tabular}
}
\caption{Evaluating different methodologies involving LLMs on requests from SPA and Education datasets. We report the  F1-score for appropriate ($F1_{\text{A}}$) and inappropriate ($F1_{\text{I}}$) class, and the total number of appropriateness judgments (Sup) for \textit{ICL}, \textit{ICL (w Undet)}, and \method{}. We boldface the best $F1_{\text{A}}$ and $F1_{\text{I}}$ scores, respectively, across all methods for each model and dataset. The evaluations show that \method{} outperforms all other methodologies for all models on both datasets.}
\label{tbl:priv_pref_eval}
\end{table*}
\begin{table}[t!]
\resizebox{\linewidth}{!}{%
\begin{tabular}{l | ccc ccc}
\toprule
\textbf{Model}   & \multicolumn{3}{c}{\textbf{Not Entailed}} & \multicolumn{3}{c}{\textbf{Entailed}} \\
 \cmidrule(lr){2-4}  \cmidrule(lr){5-7}
 & $F1_{\text{A}}$ (\%) & $F1_{\text{I}}$ (\%) & Sup & $F1_{\text{A}}$ (\%) & $F1_{\text{I}}$ (\%) & Sup \\
\midrule
\multicolumn{7}{c}{\textbf{SPA Dataset}} \\
\midrule
 Gemini 2.5 Pro & 69.6 & 79.4 & 1203 & 87.0 & 94.2 & 3797 \\
 GPT-5 & 60.4 & 79.2 & 1190 & 85.7 & 94.0 & 3810 \\
 Claude Sonnet 4.5 & 68.1 & 79.8 & 1176 & 86.5 & 94.0 & 3824 \\
 \midrule
 \multicolumn{7}{c}{\textbf{Education}} \\
 \midrule
 Gemini 2.5 Pro & 58.6 & 74.2 & 1339 & 78.7 & 86.1 & 1681 \\
 GPT-5 & 65.3 & 76.4 & 1417 & 82.0 & 88.6 & 1603 \\
 Claude Sonnet 4.5 & 68.3 & 76.9 & 1412 & 81.8 & 87.9 & 1608 \\
\bottomrule
\end{tabular}
}
\caption{Analyzing the judgments from \textit{ICL} on two sets of incoming requests: one where none of the prior requests were entailed by \method{} (Not entailed), and the other where entailment occurred by \method{} (Entailed). These results suggest that \method{} accurately identifies when the appropriateness of incoming requests are difficult to judge based on the user's prior privacy judgments.}
\label{tbl:undetermined_eval}
\end{table}

We present experiments, ablation studies, and error analysis to study the efficacy and robustness of \method{}. First, we evaluate the end-to-end performance improvement of \method{} over existing baselines in Section \ref{sec:takeaways}. Then, we evaluate how varying the number of prior user judgments affects \method{} in Section \ref{sec:ablation}. Finally, we qualitatively analyze the failure modes of \method{} in Section \ref{sec:error_analysis}.

\subsection{End-to-End Evaluation}\label{sec:takeaways}
We show the results for all approaches in Table \ref{tbl:priv_pref_eval}. Based on these results, we make the following observations:

(1) \textit{\method{} enhances personalized privacy judgments compared to pure LLM reasoning.} \method{} consistently outperforms LLM-based baselines across all datasets and reasoning models, with notable gains in appropriate judgments. For example, \method{} improves upon \textit{ICL (w Undet)} by \textbf{10.4\%} $F1_{\text{A}}$ with GPT-5 on the Education dataset. In other words, the $F1_{\text{A}}$ error reduces by \textbf{40.6\%}, demonstrating that logical entailment effectively grounds judgments in prior user privacy judgments. Notably, these improvements are magnified in smaller models. With Gemma 3 4B IT, \method{} yields $F1_{\text{A}}$ gains of 13\% and 17\% over \textit{ICL (w Undet)} on the SPA and Education datasets, respectively. This demonstrates \method{} is suitable for on-device deployment of personalized agents. \\

(2) \textit{\method{} accurately identifies when the appropriateness of an incoming request is challenging to judge.} We observe that the number of judgments that were undetermined by \method{} is higher than \textit{ICL (w Undet)}. For Gemini 2.5 Pro, we observe that \method{} judged 1203 and 1339 incoming requests as undetermined for SPA and Education datasets, respectively, compared to just 387 and 695 for \textit{ICL (w Undet)}. This observation raises the question of whether the appropriateness label of the incoming requests left undetermined by \method{} can actually be inferred by an LLM based on the user's prior privacy judgments. 

To investigate this question, we analyzed the \textit{ICL} performance on two subsets of incoming requests: those where \method{} found logical entailment, and those where it did not (see Table \ref{tbl:undetermined_eval}). We observe that ICL achieves at least a 10\% increase in both $F1_{\text{A}}$ and $F1_{\text{I}}$ on the entailed requests compared to the non-entailed ones. As such, \method{} accurately identifies requests that are answerable based on prior judgments, effectively escalating only the more challenging ones. Conversely, \textit{ICL (w Undet)} produces fewer 'undetermined' outputs, suggesting that LLMs struggle to recognize when they lack sufficient information to make a privacy judgment. This observation aligns with findings in other domains regarding LLMs' inability to abstain when lacking necessary knowledge \cite{feng2024don, kirichenko2025abstentionbench}. \\

(3) \textit{Relying only on the LLM's awareness of privacy norms is insufficient for personalization.} We observe that performance on appropriate judgments from \textit{zero-shot} is low for both models and datasets. For inappropriate judgments, \textit{zero-shot} prompting can perform much better than on appropriate judgments, but the results remain suboptimal. Hence, the LLM's awareness of privacy norms cannot sufficiently capture the nuanced behavior of user privacy preferences within the SPA and Education contexts. \\

(4) \textit{General privacy norms extracted from overall user responses for each dataset can improve personalized privacy judgments.} When comparing \textit{privacy norms} and \textit{zero-shot}, we observe that providing privacy norms to the LLMs can improve appropriate judgments for the SPA dataset across all models, and improve inappropriate judgments for the Education dataset on most models. However, including general norms can sometimes degrade performance. For instance, with Gemini 2.5 Pro on the Education dataset, $F1_{\text{A}}$ decreases by $22\%$ while $F1_{\text{I}}$ improves by only $10\%$, suggesting that adding general privacy norms is not always beneficial. \\

(5) \textit{Prior user privacy judgments improve personalized privacy judgments over general privacy norms.} By comparing the $F1_{\text{A}}$ and $F1_{\text{I}}$ between \textit{ICL} and both \textit{zero-shot} and \textit{privacy norms} for all datasets and models, it is clear that personalizing LLMs with prior user privacy judgments improves appropriate and inappropriate judgments over general privacy norms. In some cases, the improvement over general privacy norms can be as large as \textbf{50\%} for $F1_{\text{A}}$. \\

(6) \textit{Personalized privacy judgments improve when including undetermined as an option.} When comparing \textit{ICL} and \textit{ICL (w Undet)}, we observe an improvement in $F1_{\text{A}}$ and $F1_{\text{I}}$ for all LLMs. These results suggest that giving an LLM the option to abstain reduces the likelihood that the model forces an erroneous appropriateness judgment. 

\subsection{Ablation Evaluation}
\label{sec:ablation}
We evaluate how the number of prior requests affects the performance of \method{} and baseline approaches. Moreover, we investigate how the choice of reasoning affects LLM privacy judgments.

\begin{figure*}[t!]
    \centering
    \includegraphics[width=0.9\linewidth]{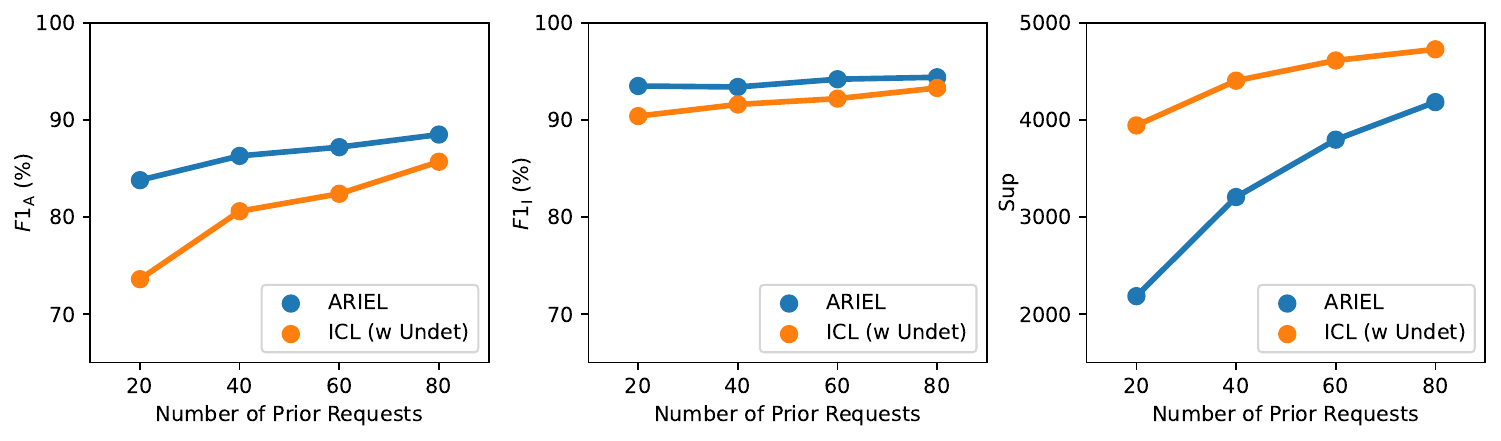}
    \caption{Ablation study on the number of prior requests with user judgments contained in user memory $D_u$. We evaluate \textit{ICL (w Undet)} and \method{} on the SPA dataset with Gemini 2.5 Pro. We report the F1-score for appropriate $F1_{\text{A}}$ and inappropriate $F1_{\text{I}}$ class, and the total number of appropriateness judgments (Sup). We find that \method{} is more robust to varying the number of prior requests compared to \textit{ICL w (Undet)}.}
    \label{fig:prior_ablation}
\end{figure*}

\begin{figure*}[t!]
    \centering
    \includegraphics[width=\linewidth]{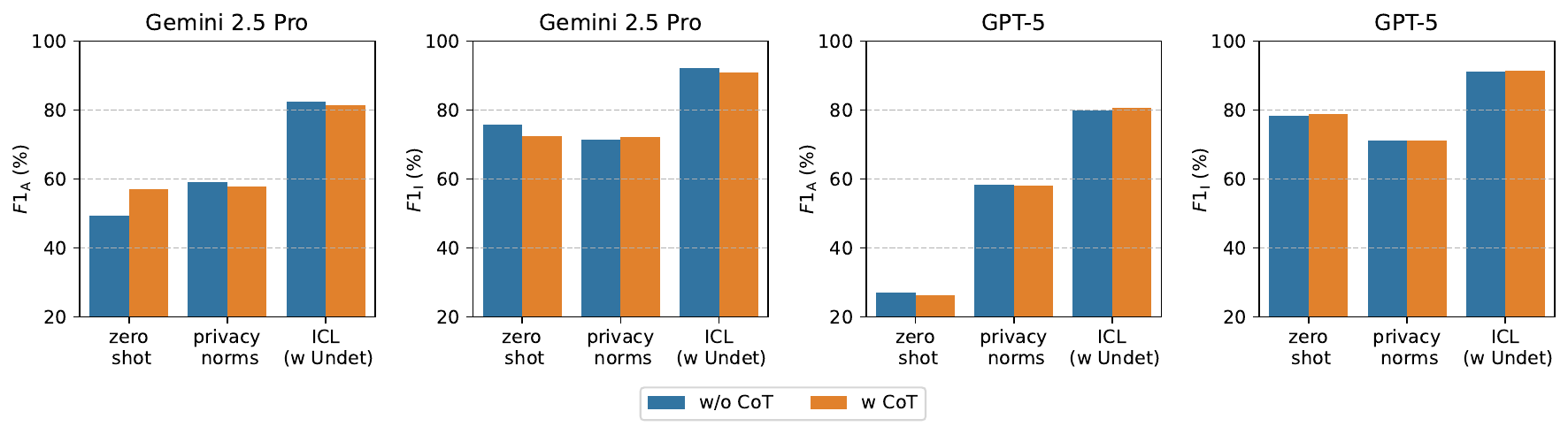}
    \caption{Ablation study on CoT. We evaluate \textit{zero-shot}, \textit{privacy norms}, and \textit{ICL w (Undet)} on the SPA dataset with Gemini 2.5 Pro. We report the F1-score for the appropriate $F1_{\text{A}}$ and inappropriate $F1_{\text{I}}$ class. The results suggest that generally CoT has negligible impact on privacy judgments across both models.}
    \label{fig:cot_ablation}
\end{figure*}

\subsubsection{Number of prior requests} We explore how varying the number of prior requests with user judgments affects the agent's judgments. Specifically, we compare the $F1_{A}$, $F1_{I}$, and Support (Sup) of \method{} against \textit{ICL (w Undet)} on the SPA dataset using Gemini 2.5 Pro. As evident in Figure \ref{fig:prior_ablation}, while the $F1_{I}$ gap remains relatively constant, the disparity in $F1_{A}$ widens significantly as the number of prior requests decreases. Notably, when reducing the prior requests from 60 to 20, $F1_{A}$ drops by only 3.4\% for \method{}, compared to an 8.8\% decline for \textit{ICL (w Undet)}. This demonstrates that \method{} is more robust to limited prior privacy judgments. Nonetheless, we observe that the number of entailments generated by \method{} decreases substantially as the number of prior requests decreases. This result is expected, as there are fewer candidate requests available to satisfy the entailment criteria (i.e., fewer prior requests differ by exactly one parameter from the incoming request).

\subsubsection{Assessing the impact of reasoning with Chain of Thought (CoT)} To assess the impact of reasoning strategy choice on appropriateness judgments, we implement CoT by prompting the LLM to provide a reasoning before making a judgment \cite{wei2022chain, yang2023large}. Inspired by findings that CoT affects privacy leakage \cite{mireshghallah2024can}, we evaluate its effect on judgment accuracy across \textit{zero-shot}, \textit{privacy norms}, and \textit{ICL (w Undet)}. Figure \ref{fig:cot_ablation} indicates that CoT typically provides negligible changes in performance. However, we observe for \textit{zero-shot} with Gemini 2.5 Pro that the $F1_{A}(\%)$ and $F1_{I}(\%)$ are significantly impacted by CoT, both positively and negatively.

While CoT provides mixed results for general privacy norms, the results clearly demonstrate that adding CoT to both \textit{ICL} and \textit{ICL (w Undet)} consistently hurts their performance. When manually inspecting the generated reasoning traces, we notice that the LLM makes misaligned assertions about the user's privacy judgments, which propagate all the way to the model's final judgment (see Section \ref{sec:error_analysis} for more information). This result also demonstrates that eliciting explanations from LLM-based reasoning can degrade appropriateness judgments. In contrast, \method{} provides a clear, traceable reasoning while maintaining well-grounded personalized privacy judgments. 

\subsection{Qualitative Analysis}
\label{sec:error_analysis}
Finally, we qualitatively analyze incorrect judgments made by both \textit{ICL (w Undet) + CoT} and \method{}. We sample 20 examples from each method on the SPA dataset with Gemini 2.5 Pro. The results are detailed in Table \ref{tbl:error_analysis}.

\subsubsection{Error Analysis on Appropriateness Judgments for \textit{ICL (w Undet) + CoT}}
Analysis of the reasoning traces reveals that Gemini 2.5 Pro occasionally performs implicit entailment when identifying prior requests that share identical parameters except for one. This behavior is noteworthy, as we did not explicitly instruct the model to use this logic, but only to extract patterns from user privacy judgments. We find that the majority of errors stem from \textit{preference misalignment} (14 examples), where the model draws incorrect equivalencies between distinct parameter values to perform this entailment. For instance, given prior requests involving "close friends" or "parents" as recipients and an incoming request involving "housemates," the model infers that the privacy judgment should remain consistent due to perceived similarity in recipient type. However, the user may not view these three recipients as similar.

We also attribute a subset of errors to the model's inherent conservatism regarding information sharing, which we describe as \textit{default to inappropriate} (5 examples). This occurs when the model concludes it cannot use the most similar prior requests that are judged as "appropriate" by the user. However, rather than outputting "undetermined", the model reasons that the judgment is inappropriate. Finally, we identified one instance of \textit{contradictory prior requests}, where the provided user history contained judgments that logically conflicted with the ground truth for the incoming request. 

\begin{table*}[t!]
\resizebox{\linewidth}{!}{%

\begin{tabular}{p{4cm}p{15cm}c}
\toprule
\textbf{Type} & \textbf{Example/Description} & \textbf{Total} \\
\midrule
\multicolumn{2}{c}{\textit{ICL (w Undet) + CoT}} \\
\midrule
 Preference misalignment & The user has previously approved sharing contacts with close friends and parents for the purpose of looking up contact details. Since housemates are a similar type of recipient, this request is likely to be considered appropriate & 14 \\
 Default to inappropriate & The user has consistently approved sharing email content with advertising agencies for personalized ads, as long as there are some safeguards in place. Since this request has no safeguards, it is inappropriate. & 5 \\
 Contradictory Prior Requests & The user has previously judged sharing bank account details with business and finance skills for making transactions as inappropriate without any safeguards. Since the current request lacks a transmission principle, it is likely to be considered inappropriate. & 1 \\
 
 \midrule
 \multicolumn{2}{c}{\method{}} \\
 \midrule
 
 Ontology Generation & An error caused by the hierarchy of the ontology being incorrect. This is due to incomplete information from the user's privacy preferences. For example, "housemate" only appears once in the prior request which is not enough knowledge to determine how trustworthy housemate is for the user. & 10 \\
 Ontology Mapping & This typically caused by domain specific issues about the dataset. For SPA, the transmission principle has many combinations for puprose and condition, which can make it hard for the model to map to the correct level. & 5 \\
 Contradictory Prior Requests & The judgments of the prior requests contradict the judgment of the incoming request. & 5 \\
\bottomrule
\end{tabular}
}
\caption{The different error types of 20 judgments from \textit{ICL (w Undet) + CoT} and \method{} on SPA dataset with Gemini 2.5 Pro.}
\label{tbl:error_analysis}
\end{table*}

\subsubsection{Error Analysis on Appropriateness Judgments for \method{}}
We classify errors in \method{} into three categories: (1) ontology generation, (2) ontology mapping, and (3) contradictory prior requests. The majority of errors stem from ontology generation, where the generated ontology produces incorrect entailments (10 examples). This typically arises from data sparsity when specific parameter values (e.g., 'housemates' or 'close friends') appear only once or twice in the user's prior requests. The model struggles to accurately reflect the user's privacy preferences onto the ontology. A second source of error is ontology mapping, where the model fails to map parameter values to the ontology correctly (5 examples). This is often caused by domain-specific idiosyncrasies in the evaluation dataset. For instance, the transmission principle has a specific format combining both purpose and condition, which can be challenging for the model to map. Finally, the remaining errors (5 examples) result from contradictory prior requests. For instance, there was a prior request where the user had judged sharing bank account details with business and finance skills for the purpose of making transactions as inappropriate. The incoming request is identical to the prior, except there is no purpose. Thus, the incoming request should also be inappropriate, since removing the purpose likely makes the request more inappropriate. Both \textit{ICL (w Undet) + CoT} and \method{} judged this request inappropriate; however, the user judged the incoming request to be appropriate.

Potential solutions for (1) involve providing more user judgments on prior requests during ontology generation, while solutions for (2) involve providing more information on the domain-specific format of each parameter in the ontology mapping prompt. The errors in (3) represent an open challenge that is discussed in more detail in Section~\ref{sec:personalization_challenges}.
\section{Discussion}\label{sec:discussion}
Based on our results and takeaways, we discuss in more detail the broad implications of this work as well as its limitations.

\subsection{Personalization Challenges}\label{sec:personalization_challenges}
In this paper, we consider the user's judgments on prior data-sharing requests as proxies for the user's privacy preferences \cite{wijesekera2017feasibility, colnago2022concern}. Personalizing privacy decisions based on these preferences introduces several challenges that must be considered. 

\paragraph{Contradictory Privacy Preferences} 
Highlighted by the error analysis in Section \ref{sec:error_analysis}, contradictions can arise within the user's own history of prior judgments. For example, the user says it is inappropriate to share their banking information with a third party if their data was confidential, but deems it appropriate if there is no safeguard on the data. In such scenarios, the agent may execute the entailment logic correctly but still reach an incorrect privacy judgment. Consequently, the agent must (1) detect these user privacy judgment contradictions and (2) resolve them, typically by escalating the request to the user. The fundamental difficulty lies in distinguishing whether an incorrect privacy judgment stems from an error in the entailment process of \method{} or an inconsistency in the user's prior privacy judgments.

\paragraph{Evolving Privacy Preferences} 
The second challenge involves the dynamic nature of privacy preferences as they can change over time. For example, a user can become more conservative about sharing their data as they get older. Currently, \method{} does not support automatic detection of privacy preference changes. Hence, future adaptations of \method{} require a mechanism to detect when the user's privacy preferences are shifting and determine which preferences must be updated. One potential mechanism is to periodically escalate incoming requests to the user to confirm the agent's privacy decision. 

\subsection{User Agency in Personal AI Agents}
While personalized agents reduce the burden of privacy management, they must not remove user agency \cite{mitchell2025fully, zhang2025autonomy}. This observation motivated the entailment logic of \method{}, where the agent automates clear decisions that can be entailed but escalates requests where the judgment cannot be easily inferred by the user's prior judgments. Our results confirm that identifying these non-entailed incoming requests drives \method{}'s judgment accuracy gains. Furthermore, by providing transparent reasoning traces for user auditing, \method{} successfully balances automation with necessary human oversight.

Although \method{} was designed with user agency as a core requirement, further work is needed to strengthen the interaction between the user and the agent. One promising direction is the development of a dedicated user interface (UI) that enables users to: (1) visualize the reasoning steps behind \method{}'s decisions (e.g., which requests were entailed and how they were mapped in the ontology), and (2) intervene in the decision-making process when necessary. Subsequently, this UI can facilitate user studies to evaluate how well \method{}'s privacy reasoning and judgments align with the user's own privacy reasoning \cite{nezhad2026understanding}. Furthermore, this UI could be used to evaluate potential solutions for the challenges outlined in Section~\ref{sec:personalization_challenges}, such as user verification of \method{}'s reasoning and decisions to identify contradictions or changes in the user's privacy preferences.

\subsection{Personalization and Privacy Norms}
Our results in Section~\ref{sec:user_pref_eval} suggest that effective personalization of LLM agents requires grounding judgments in prior user privacy judgments, as general norms are insufficient. However, this finding does not invalidate alignment of LLMs with general contextual norms, as this alignment remains crucial for many data-sharing scenarios \cite{mireshghallah2024can, shao2024privacylens}. Our work highlights that when personalization is a crucial component for making privacy decisions, then the LLM's privacy judgments should be grounded in user-specific preferences rather than general privacy norms.

\subsection{Limitations}\label{sec:limitations}
We identify four main limitations involving our work. 

\paragraph{Ontology Mapping and Generation} 
\method{} generates generalizable ontologies for individual fields (e.g., data type) based on the user's prior privacy judgments. However, in practice, the ontology of one field may depend on the values of others. For instance, data sensitivity can vary based on the recipient: a user may consider their income more sensitive than SSN when sharing with a parent, but the sensitivity is reversed when sharing with a friend. Our results in Section 6.1 demonstrate that even without explicitly modeling these conditional dependencies, \method{} accurately infers privacy judgments. We leave the incorporation of context-conditional hierarchies for future work.

\paragraph{Evaluation Datasets} 
Our evaluation datasets simulate prior requests by partitioning the user's judgments on requests into two pairwise disjoint sets, where one set stores already observed requests while the other contains new requests. While we believe this reasonably approximates our problem setup, it does not incorporate user decision-making over time. Unfortunately, to our knowledge, there are no existing publicly-available datasets that incorporate user privacy preferences over time. An interesting direction for future work could elicit user privacy preferences over a period of time, rather than giving a one-time questionnaire to participants. 

Although we believe our evaluation datasets accurately reflect our problem setup, they were not originally framed as AI agents acting on the users' behalf. This distinction is notable, as prior work indicates that privacy preferences can shift when information-sharing decisions are delegated to a personal AI agent \cite{guo2025not}. While recent studies have begun eliciting user privacy preferences regarding AI agents and chatbots \cite{guo2025not, tran2025understanding, zhang2025autonomy, liu2025prevalence, wu2025towards, zufferey2025ai}, these datasets are either (1) not publicly available, or (2) restricted to narrowly scoped scenarios. Consequently, this emerging field would benefit significantly from future work that appropriately releases open datasets on user privacy preferences in agentic data-sharing contexts.

\paragraph{Cold-start Cases} 
Our evaluation does not consider cold-start cases where the user has not made prior judgments on requests-- essentially, no prior privacy preferences. Handling these cold-start cases is an important consideration when designing personalized agents. While falling back to general privacy norms is a potential solution, our results demonstrate they are often insufficient for personalized privacy decisions. An alternative solution could be to develop the minimal set of questions to elicit the user's privacy preferences \cite{guo2025privi} while offsetting user burden. We leave evaluation on cold-start cases as future work.  

\paragraph{Non-adversarial Setting} 
Lastly, our work assumes a non-adversarial setting which could limit the applicability of \method{} in open environments. However, \method{} is designed as a specialized module focused on agent privacy decisions, functioning within a larger system that authenticates inputs. An example is Google Play Store where app developers are verified and users provide feedback on problematic apps, which prevents requests to the agent from being maliciously designed. Thus, defense against adversarial attacks, such as prompt injection \cite{greshake2023not}, is delegated to separate system components that focus specifically on these attacks \cite{meng2025cellmate, tsai2025contextual}. Additionally, ARIEL operates on limited structured input containing just five fields, inherently reducing the risk of adversarial attacks. 
\section{Conclusion}
In this work, we explored personalization of LLM-based agents for making privacy judgments. We identified that general privacy norms are insufficient for effective personalization, and LLM-based reasoning with ICL does not guarantee that judgments are faithfully grounded in the user's prior privacy judgments. We addressed these limitations by designing \method{} to ensure judgments are grounded in the user's prior judgments with clear and traceable reasoning that the user can reference. Our evaluations demonstrate that \method{} can reduce the F1 score error for appropriate judgments by up to \textbf{40.6\%} over LLM reasoning. Overall, our findings suggest that jointly leveraging LLMs with logical entailment is a promising direction for designing agentic frameworks for personalized privacy judgments. 

\begin{acks}
    We would like to thank Adrià Gascón, Sarah Meiklejohn, Eugene Bagdasarian, Daniel Ramage, Zhichen Zhao, and Ku Kogan for their constructive discussions and support. The authors used generative AI-based tools to revise the text, improve flow and correct any typos, grammatical errors, and awkward phrasing. This work is supported by NSF award number DGE-1842487 (J.F.) and the REAL@USC-Meta center (J.F., M.A.).
\end{acks}

\bibliographystyle{ACM-Reference-Format}
\bibliography{references}

@article{shao2024privacylens,
  title={Privacylens: Evaluating privacy norm awareness of language models in action},
  author={Shao, Yijia and Li, Tianshi and Shi, Weiyan and Liu, Yanchen and Yang, Diyi},
  journal={Advances in Neural Information Processing Systems},
  volume={37},
  pages={89373--89407},
  year={2024}
}

@inproceedings{mireshghallah2024can,
  title={Can LLMs Keep a Secret? Testing Privacy Implications of Language Models via Contextual Integrity Theory},
  author={Mireshghallah, Niloofar and Kim, Hyunwoo and Zhou, Xuhui and Tsvetkov, Yulia and Sap, Maarten and Shokri, Reza and Choi, Yejin},
  booktitle={ICLR},
  year={2024}
}

@article{choi2018role,
  title={The role of privacy fatigue in online privacy behavior},
  author={Choi, Hanbyul and Park, Jonghwa and Jung, Yoonhyuk},
  journal={Computers in Human Behavior},
  volume={81},
  pages={42--51},
  year={2018},
  publisher={Elsevier}
}

@article{nissenbaum2004privacy,
  title={Privacy as contextual integrity},
  author={Nissenbaum, Helen},
  journal={Wash. L. Rev.},
  volume={79},
  pages={119},
  year={2004},
  publisher={HeinOnline}
}

@inproceedings{zhan2023privacy,
  title={Privacy-enhanced personal assistants based on dialogues and case similarity},
  author={Zhan, Nicole and Sarkadi, Stefan and Such, Jose},
  booktitle={European Conference on Artificial Intelligence},
  year={2023},
  organization={IOS Press}
}

@inproceedings{zhan2022model,
  title={A model for governing information sharing in smart assistants},
  author={Zhan, Xiao and Sarkadi, Stefan and Criado, Natalia and Such, Jose},
  booktitle={Proceedings of the 2022 AAAI/ACM Conference on AI, Ethics, and Society},
  pages={845--855},
  year={2022}
}

@inproceedings{amoros2023predicting,
  title={Predicting privacy preferences for smart devices as norms},
  author={Amoros, Marc Serramia and Seymour, William and Criado, Natalia and Luck, Michael},
  booktitle={The 22nd International Conference on Autonomous Agents and Multiagent Systems},
  year={2023},
  organization={International Foundation for Autonomous Agents and Multiagent Systems (IFAAMAS)}
}

@inproceedings{barkhuus2012mismeasurement,
  title={The mismeasurement of privacy: using contextual integrity to reconsider privacy in HCI},
  author={Barkhuus, Louise},
  booktitle={Proceedings of the SIGCHI Conference on Human Factors in Computing Systems},
  pages={367--376},
  year={2012}
}

@inproceedings{abdi2021privacy,
  title={Privacy norms for smart home personal assistants},
  author={Abdi, Noura and Zhan, Xiao and Ramokapane, Kopo M and Such, Jose},
  booktitle={Proceedings of the 2021 CHI conference on human factors in computing systems},
  pages={1--14},
  year={2021}
}

@inproceedings{kokciyan2022taking,
  title={Taking Situation-Based Privacy Decisions: Privacy Assistants Working with Humans.},
  author={K{\"o}kciyan, Nadin and Yolum, Pinar and others},
  booktitle={IJCAI},
  pages={703--709},
  year={2022}
}

@article{wang2024survey,
  title={A survey on large language model based autonomous agents},
  author={Wang, Lei and Ma, Chen and Feng, Xueyang and Zhang, Zeyu and Yang, Hao and Zhang, Jingsen and Chen, Zhiyuan and Tang, Jiakai and Chen, Xu and Lin, Yankai and others},
  journal={Frontiers of Computer Science},
  volume={18},
  number={6},
  pages={186345},
  year={2024},
  publisher={Springer}
}

@article{brown2020language,
  title={Language models are few-shot learners},
  author={Brown, Tom and Mann, Benjamin and Ryder, Nick and Subbiah, Melanie and Kaplan, Jared D and Dhariwal, Prafulla and Neelakantan, Arvind and Shyam, Pranav and Sastry, Girish and Askell, Amanda and others},
  journal={Advances in neural information processing systems},
  volume={33},
  pages={1877--1901},
  year={2020}
}

@article{morel2025ai,
  title={AI-driven Personalized Privacy Assistants: a Systematic Literature Review},
  author={Morel, Victor and Iwaya, Leonardo Horn and Fischer-H{\"u}bner, Simone},
  journal={IEEE Access},
  year={2025},
  publisher={IEEE}
}

@inproceedings{liu2016follow,
  title={Follow my recommendations: A personalized privacy assistant for mobile app permissions},
  author={Liu, Bin and Andersen, Mads Schaarup and Schaub, Florian and Almuhimedi, Hazim and Zhang, Shikun Aerin and Sadeh, Norman and Agarwal, Yuvraj and Acquisti, Alessandro},
  booktitle={Twelfth symposium on usable privacy and security (SOUPS 2016)},
  pages={27--41},
  year={2016}
}

@inproceedings{wijesekera2018contextualizing,
  title={Contextualizing privacy decisions for better prediction (and protection)},
  author={Wijesekera, Primal and Reardon, Joel and Reyes, Irwin and Tsai, Lynn and Chen, Jung-Wei and Good, Nathan and Wagner, David and Beznosov, Konstantin and Egelman, Serge},
  booktitle={Proceedings of the 2018 CHI Conference on Human Factors in Computing Systems},
  pages={1--13},
  year={2018}
}

@article{stover2023investigating,
  title={Investigating how users imagine their personal privacy assistant},
  author={St{\"o}ver, Alina and Hahn, Sara and Kretschmer, Felix and Gerber, Nina},
  journal={Proceedings on Privacy Enhancing Technologies},
  year={2023}
}

@inproceedings{naeini2017privacy,
  title={Privacy expectations and preferences in an $\{$IoT$\}$ world},
  author={Naeini, Pardis Emami and Bhagavatula, Sruti and Habib, Hana and Degeling, Martin and Bauer, Lujo and Cranor, Lorrie Faith and Sadeh, Norman},
  booktitle={Thirteenth symposium on usable privacy and security (SOUPS 2017)},
  pages={399--412},
  year={2017}
}

@inproceedings{lin2014modeling,
  title={Modeling $\{$Users’$\}$ mobile app privacy preferences: Restoring usability in a sea of permission settings},
  author={Lin, Jialiu and Liu, Bin and Sadeh, Norman and Hong, Jason I},
  booktitle={10th Symposium On Usable Privacy and Security (SOUPS 2014)},
  pages={199--212},
  year={2014}
}

@article{barbosa2019if,
  title={“What if?” Predicting individual users’ smart home privacy preferences and their changes},
  author={Barbosa, Nat{\~a} M and Park, Joon S and Yao, Yaxing and Wang, Yang},
  journal={Proceedings on Privacy Enhancing Technologies},
  year={2019}
}

@inproceedings{shvartzshnaider2016learning,
  title={Learning privacy expectations by crowdsourcing contextual informational norms},
  author={Shvartzshnaider, Yan and Tong, Schrasing and Wies, Thomas and Kift, Paula and Nissenbaum, Helen and Subramanian, Lakshminarayanan and Mittal, Prateek},
  booktitle={Proceedings of the AAAI Conference on Human Computation and Crowdsourcing},
  volume={4},
  pages={209--218},
  year={2016}
}

@article{apthorpe2018discovering,
  title={Discovering smart home internet of things privacy norms using contextual integrity},
  author={Apthorpe, Noah and Shvartzshnaider, Yan and Mathur, Arunesh and Reisman, Dillon and Feamster, Nick},
  journal={Proceedings of the ACM on interactive, mobile, wearable and ubiquitous technologies},
  volume={2},
  number={2},
  pages={1--23},
  year={2018},
  publisher={ACM New York, NY, USA}
}

@inproceedings{apthorpe2019evaluating,
  title={Evaluating the Contextual Integrity of Privacy Regulation: Parents'$\{$IoT$\}$ Toy Privacy Norms Versus $\{$COPPA$\}$},
  author={Apthorpe, Noah and Varghese, Sarah and Feamster, Nick},
  booktitle={28th USENIX security symposium (USENIX security 19)},
  pages={123--140},
  year={2019}
}

@incollection{nissenbaum2009privacy,
  title={Privacy in context: Technology, policy, and the integrity of social life},
  author={Nissenbaum, Helen},
  booktitle={Privacy in context},
  year={2009},
  publisher={Stanford University Press}
}

@article{bhatia2018empirical,
  title={Empirical measurement of perceived privacy risk},
  author={Bhatia, Jaspreet and Breaux, Travis D},
  journal={ACM Transactions on Computer-Human Interaction (TOCHI)},
  volume={25},
  number={6},
  pages={1--47},
  year={2018},
  publisher={ACM New York, NY, USA}
}

@article{hoyle2020privacy,
  title={Privacy norms and preferences for photos posted online},
  author={Hoyle, Roberto and Stark, Luke and Ismail, Qatrunnada and Crandall, David and Kapadia, Apu and Anthony, Denise},
  journal={ACM Transactions on Computer-Human Interaction (TOCHI)},
  volume={27},
  number={4},
  pages={1--27},
  year={2020},
  publisher={ACM New York, NY, USA}
}

@article{martin2016measuring,
  title={Measuring privacy: An empirical test using context to expose confounding variables},
  author={Martin, Kirsten and Nissenbaum, Helen},
  journal={Colum. Sci. \& Tech. L. Rev.},
  volume={18},
  pages={176},
  year={2016},
  publisher={HeinOnline}
}

@article{guo2025not,
  title={Not my agent, not my boundary? elicitation of personal privacy boundaries in ai-delegated information sharing},
  author={Guo, Bingcan and Xu, Eryue and Zhang, Zhiping and Li, Tianshi},
  journal={arXiv preprint arXiv:2509.21712},
  year={2025}
}

@article{li2025privaci,
  title={Privaci-bench: Evaluating privacy with contextual integrity and legal compliance},
  author={Li, Haoran and Hu, Wenbin and Jing, Huihao and Chen, Yulin and Hu, Qi and Han, Sirui and Chu, Tianshu and Hu, Peizhao and Song, Yangqiu},
  journal={arXiv preprint arXiv:2502.17041},
  year={2025}
}

@article{yi2025privacy,
  title={Privacy Reasoning in Ambiguous Contexts},
  author={Yi, Ren and Suciu, Octavian and Gascon, Adria and Meiklejohn, Sarah and Bagdasarian, Eugene and Gruteser, Marco},
  journal={arXiv preprint arXiv:2506.12241},
  year={2025}
}

@article{sullivan2025benchmarking,
  title={Benchmarking LLM Privacy Recognition for Social Robot Decision Making},
  author={Sullivan, Dakota and Zhang, Shirley and Li, Jennica and Kirkorian, Heather and Mutlu, Bilge and Fawaz, Kassem},
  journal={arXiv preprint arXiv:2507.16124},
  year={2025}
}

@article{lan2025contextual,
  title={Contextual integrity in llms via reasoning and reinforcement learning},
  author={Lan, Guangchen and Inan, Huseyin A and Abdelnabi, Sahar and Kulkarni, Janardhan and Wutschitz, Lukas and Shokri, Reza and Brinton, Christopher G and Sim, Robert},
  journal={arXiv preprint arXiv:2506.04245},
  year={2025}
}

@article{wei2022chain,
  title={Chain-of-thought prompting elicits reasoning in large language models},
  author={Wei, Jason and Wang, Xuezhi and Schuurmans, Dale and Bosma, Maarten and Xia, Fei and Chi, Ed and Le, Quoc V and Zhou, Denny and others},
  journal={Advances in neural information processing systems},
  volume={35},
  pages={24824--24837},
  year={2022}
}

@article{cheng2024ci,
  title={Ci-bench: Benchmarking contextual integrity of ai assistants on synthetic data},
  author={Cheng, Zhao and Wan, Diane and Abueg, Matthew and Ghalebikesabi, Sahra and Yi, Ren and Bagdasarian, Eugene and Balle, Borja and Mellem, Stefan and O'Banion, Shawn},
  journal={arXiv preprint arXiv:2409.13903},
  year={2024}
}

@article{ngong2025protecting,
  title={Protecting users from themselves: Safeguarding contextual privacy in interactions with conversational agents},
  author={Ngong, Ivoline and Kadhe, Swanand and Wang, Hao and Murugesan, Keerthiram and Weisz, Justin D and Dhurandhar, Amit and Ramamurthy, Karthikeyan Natesan},
  journal={arXiv preprint arXiv:2502.18509},
  year={2025}
}

@article{ghalebikesabi2024operationalizing,
  title={Operationalizing contextual integrity in privacy-conscious assistants},
  author={Ghalebikesabi, Sahra and Bagdasaryan, Eugene and Yi, Ren and Yona, Itay and Shumailov, Ilia and Pappu, Aneesh and Shi, Chongyang and Weidinger, Laura and Stanforth, Robert and Berrada, Leonard and others},
  journal={arXiv preprint arXiv:2408.02373},
  year={2024}
}

@inproceedings{bagdasarian2024airgapagent,
  title={Airgapagent: Protecting privacy-conscious conversational agents},
  author={Bagdasarian, Eugene and Yi, Ren and Ghalebikesabi, Sahra and Kairouz, Peter and Gruteser, Marco and Oh, Sewoong and Balle, Borja and Ramage, Daniel},
  booktitle={Proceedings of the 2024 on ACM SIGSAC Conference on Computer and Communications Security},
  pages={3868--3882},
  year={2024}
}

@article{zhang2023siren,
  title={Siren's Song in the AI Ocean: A Survey on Hallucination in Large Language Models},
  author={Zhang, Yue and Li, Yafu and Cui, Leyang and Cai, Deng and Liu, Lemao and Fu, Tingchen and Huang, Xinting and Zhao, Enbo and Zhang, Yu and Chen, Yulong and others},
  journal={arXiv e-prints},
  pages={arXiv--2309},
  year={2023}
}

@article{ji2023survey,
  title={Survey of hallucination in natural language generation},
  author={Ji, Ziwei and Lee, Nayeon and Frieske, Rita and Yu, Tiezheng and Su, Dan and Xu, Yan and Ishii, Etsuko and Bang, Ye Jin and Madotto, Andrea and Fung, Pascale},
  journal={ACM computing surveys},
  volume={55},
  number={12},
  pages={1--38},
  year={2023},
  publisher={ACM New York, NY}
}

@article{lin2025zebralogic,
  title={Zebralogic: On the scaling limits of llms for logical reasoning},
  author={Lin, Bill Yuchen and Bras, Ronan Le and Richardson, Kyle and Sabharwal, Ashish and Poovendran, Radha and Clark, Peter and Choi, Yejin},
  journal={arXiv preprint arXiv:2502.01100},
  year={2025}
}

@article{barez2025chain,
  title={Chain-of-thought is not explainability},
  author={Barez, Fazl and Wu, Tung-Yu and Arcuschin, Iv{\'a}n and Lan, Michael and Wang, Vincent and Siegel, Noah and Collignon, Nicolas and Neo, Clement and Lee, Isabelle and Paren, Alasdair and others},
  journal={Preprint, alphaXiv},
  pages={v1},
  year={2025}
}

@article{kalai2025language,
  title={Why language models hallucinate},
  author={Kalai, Adam Tauman and Nachum, Ofir and Vempala, Santosh S and Zhang, Edwin},
  journal={arXiv preprint arXiv:2509.04664},
  year={2025}
}

@inproceedings{wang2018guileak,
  title={Guileak: Tracing privacy policy claims on user input data for android applications},
  author={Wang, Xiaoyin and Qin, Xue and Hosseini, Mitra Bokaei and Slavin, Rocky and Breaux, Travis D and Niu, Jianwei},
  booktitle={Proceedings of the 40th International Conference on Software Engineering},
  pages={37--47},
  year={2018}
}

@inproceedings{andow2019policylint,
  title={$\{$PolicyLint$\}$: investigating internal privacy policy contradictions on google play},
  author={Andow, Benjamin and Mahmud, Samin Yaseer and Wang, Wenyu and Whitaker, Justin and Enck, William and Reaves, Bradley and Singh, Kapil and Xie, Tao},
  booktitle={28th USENIX security symposium (USENIX security 19)},
  pages={585--602},
  year={2019}
}

@inproceedings{andow2020actions,
  title={Actions speak louder than words:$\{$Entity-Sensitive$\}$ privacy policy and data flow analysis with $\{$PoliCheck$\}$},
  author={Andow, Benjamin and Mahmud, Samin Yaseer and Whitaker, Justin and Enck, William and Reaves, Bradley and Singh, Kapil and Egelman, Serge},
  booktitle={29th USENIX Security Symposium (USENIX Security 20)},
  pages={985--1002},
  year={2020}
}

@article{zhang2025towards,
  title={Towards Aligning Personalized Conversational Recommendation Agents with Users' Privacy Preferences},
  author={Zhang, Shuning and Ma, Ying and Chen, Jingruo and Li, Simin and Yi, Xin and Li, Hewu},
  journal={arXiv preprint arXiv:2508.07672},
  year={2025}
}

@inproceedings{tran2025understanding,
  title={Understanding Privacy Norms Around LLM-Based Chatbots: A Contextual Integrity Perspective},
  author={Tran, Sarah and Lu, Hongfan and Slaughter, Isaac and Herman, Bernease and Dangol, Aayushi and Fu, Yue and Chen, Lufei and Gebreyohannes, Biniyam and Howe, Bill and Hiniker, Alexis and others},
  booktitle={Proceedings of the AAAI/ACM Conference on AI, Ethics, and Society},
  volume={8},
  number={3},
  pages={2522--2534},
  year={2025}
}

@article{feng2024don,
  title={Don't hallucinate, abstain: Identifying LLM knowledge gaps via multi-LLM collaboration},
  author={Feng, Shangbin and Shi, Weijia and Wang, Yike and Ding, Wenxuan and Balachandran, Vidhisha and Tsvetkov, Yulia},
  journal={arXiv preprint arXiv:2402.00367},
  year={2024}
}

@article{olausson2023linc,
  title={LINC: A neurosymbolic approach for logical reasoning by combining language models with first-order logic provers},
  author={Olausson, Theo X and Gu, Alex and Lipkin, Benjamin and Zhang, Cedegao E and Solar-Lezama, Armando and Tenenbaum, Joshua B and Levy, Roger},
  journal={arXiv preprint arXiv:2310.15164},
  year={2023}
}

@article{trinh2024solving,
  title={Solving olympiad geometry without human demonstrations},
  author={Trinh, Trieu H and Wu, Yuhuai and Le, Quoc V and He, He and Luong, Thang},
  journal={Nature},
  volume={625},
  number={7995},
  pages={476--482},
  year={2024},
  publisher={Nature Publishing Group UK London}
}

@article{pan2023logic,
  title={Logic-lm: Empowering large language models with symbolic solvers for faithful logical reasoning},
  author={Pan, Liangming and Albalak, Alon and Wang, Xinyi and Wang, William Yang},
  journal={arXiv preprint arXiv:2305.12295},
  year={2023}
}

@article{ye2022unreliability,
  title={The unreliability of explanations in few-shot prompting for textual reasoning},
  author={Ye, Xi and Durrett, Greg},
  journal={Advances in neural information processing systems},
  volume={35},
  pages={30378--30392},
  year={2022}
}

@article{ye2023satlm,
  title={Satlm: Satisfiability-aided language models using declarative prompting},
  author={Ye, Xi and Chen, Qiaochu and Dillig, Isil and Durrett, Greg},
  journal={Advances in Neural Information Processing Systems},
  volume={36},
  pages={45548--45580},
  year={2023}
}

@article{kirichenko2025abstentionbench,
  title={AbstentionBench: Reasoning LLMs Fail on Unanswerable Questions},
  author={Kirichenko, Polina and Ibrahim, Mark and Chaudhuri, Kamalika and Bell, Samuel J},
  journal={arXiv preprint arXiv:2506.09038},
  year={2025}
}

@article{team2025gemma,
  title={Gemma 3 technical report},
  author={Team, Gemma and Kamath, Aishwarya and Ferret, Johan and Pathak, Shreya and Vieillard, Nino and Merhej, Ramona and Perrin, Sarah and Matejovicova, Tatiana and Ram{\'e}, Alexandre and Rivi{\`e}re, Morgane and others},
  journal={arXiv preprint arXiv:2503.19786},
  year={2025}
}

@inproceedings{alshamsan2022detecting,
  title={Detecting privacy policies violations using natural language inference (nli)},
  author={Alshamsan, Abdullah R and Chaudhry, Shafique A},
  booktitle={2022 IEEE Asia-Pacific Conference on Computer Science and Data Engineering (CSDE)},
  pages={1--6},
  year={2022},
  organization={IEEE}
}

@inproceedings{rao2016expecting,
  title={Expecting the unexpected: Understanding mismatched privacy expectations online},
  author={Rao, Ashwini and Schaub, Florian and Sadeh, Norman and Acquisti, Alessandro and Kang, Ruogu},
  booktitle={Twelfth Symposium on Usable Privacy and Security (SOUPS 2016)},
  pages={77--96},
  year={2016}
}

@inproceedings{hosseini2021ambiguity,
  title={Ambiguity and generality in natural language privacy policies},
  author={Hosseini, Mitra Bokaei and Heaps, John and Slavin, Rocky and Niu, Jianwei and Breaux, Travis},
  booktitle={2021 IEEE 29th International Requirements Engineering Conference (RE)},
  pages={70--81},
  year={2021},
  organization={IEEE}
}

@article{groschupp2025can,
  title={Can LLMs Make (Personalized) Access Control Decisions?},
  author={Groschupp, Friederike and Lain, Daniele and Dhar, Aritra and Lazier, Lara Magdalena and {\v{C}}apkun, Srdjan},
  journal={arXiv preprint arXiv:2511.20284},
  year={2025}
}

@article{gallegos2024bias,
  title={Bias and fairness in large language models: A survey},
  author={Gallegos, Isabel O and Rossi, Ryan A and Barrow, Joe and Tanjim, Md Mehrab and Kim, Sungchul and Dernoncourt, Franck and Yu, Tong and Zhang, Ruiyi and Ahmed, Nesreen K},
  journal={Computational Linguistics},
  volume={50},
  number={3},
  pages={1097--1179},
  year={2024},
  publisher={MIT Press 255 Main Street, 9th Floor, Cambridge, Massachusetts 02142, USA~…}
}

@inproceedings{fan2024goldcoin,
  title={GoldCoin: Grounding Large Language Models in Privacy Laws via Contextual Integrity Theory},
  author={Fan, Wei and Li, Haoran and Deng, Zheye and Wang, Weiqi and Song, Yangqiu},
  booktitle={Proceedings of the 2024 Conference on Empirical Methods in Natural Language Processing},
  pages={3321--3343},
  year={2024}
}

@article{acquisti2006there,
  title={Is there a cost to privacy breaches? An event study},
  author={Acquisti, Alessandro and Friedman, Allan and Telang, Rahul},
  journal={ICIS 2006 proceedings},
  pages={94},
  year={2006}
}

@article{mitchell2025fully,
  title={Fully autonomous ai agents should not be developed},
  author={Mitchell, Margaret and Ghosh, Avijit and Luccioni, Alexandra Sasha and Pistilli, Giada},
  journal={arXiv preprint arXiv:2502.02649},
  year={2025}
}

@article{zhang2025autonomy,
  title={Autonomy Matters: A Study on Personalization-Privacy Dilemma in LLM Agents},
  author={Zhang, Zhiping and Zhang, Yi Evie and Shi, Freda and Li, Tianshi},
  journal={arXiv preprint arXiv:2510.04465},
  year={2025}
}

@inproceedings{guo2025privi,
  title={Privi: Assisting Users in Authoring Contextual Privacy Rules with an LLM Sandbox},
  author={Guo, Bingcan and Zhang, Zhiping and Li, Tianshi},
  booktitle={Proceedings of the 2025 Workshop on Human-Centered AI Privacy and Security},
  pages={73--83},
  year={2025}
}

@inproceedings{tsai2025contextual,
  title={Contextual Agent Security: A Policy for Every Purpose},
  author={Tsai, Lillian and Bagdasarian, Eugene},
  booktitle={Proceedings of the 2025 Workshop on Hot Topics in Operating Systems},
  pages={8--17},
  year={2025}
}

@article{meng2025cellmate,
  title={cellmate: Sandboxing browser ai agents},
  author={Meng, Luoxi and Feng, Henry and Shumailov, Ilia and Fernandes, Earlence},
  journal={arXiv preprint arXiv:2512.12594},
  year={2025}
}

@inproceedings{greshake2023not,
  title={Not what you've signed up for: Compromising real-world llm-integrated applications with indirect prompt injection},
  author={Greshake, Kai and Abdelnabi, Sahar and Mishra, Shailesh and Endres, Christoph and Holz, Thorsten and Fritz, Mario},
  booktitle={Proceedings of the 16th ACM workshop on artificial intelligence and security},
  pages={79--90},
  year={2023}
}

@inproceedings{wijesekera2017feasibility,
  title={The feasibility of dynamically granted permissions: Aligning mobile privacy with user preferences},
  author={Wijesekera, Primal and Baokar, Arjun and Tsai, Lynn and Reardon, Joel and Egelman, Serge and Wagner, David and Beznosov, Konstantin},
  booktitle={2017 IEEE Symposium on Security and Privacy (SP)},
  pages={1077--1093},
  year={2017},
  organization={IEEE}
}

@article{nezhad2026understanding,
  title={Understanding Users' Privacy Reasoning and Behaviors During Chatbot Use to Support Meaningful Agency in Privacy},
  author={Nezhad, Mohammad Hadi and Castro, Francisco Enrique Vicente and Arroyo, Ivon},
  journal={arXiv preprint arXiv:2601.18125},
  year={2026}
}

@article{mireshghallah2025cimemories,
  title={Cimemories: A compositional benchmark for contextual integrity of persistent memory in llms},
  author={Mireshghallah, Niloofar and Mangaokar, Neal and Kokhlikyan, Narine and Zharmagambetov, Arman and Zaheer, Manzil and Mahloujifar, Saeed and Chaudhuri, Kamalika},
  journal={arXiv preprint arXiv:2511.14937},
  year={2025}
}

@inproceedings{colnago2020informing,
  title={Informing the design of a personalized privacy assistant for the internet of things},
  author={Colnago, Jessica and Feng, Yuanyuan and Palanivel, Tharangini and Pearman, Sarah and Ung, Megan and Acquisti, Alessandro and Cranor, Lorrie Faith and Sadeh, Norman},
  booktitle={Proceedings of the 2020 CHI Conference on Human Factors in Computing Systems},
  pages={1--13},
  year={2020}
}

@inproceedings{colnago2022concern,
  title={Is it a concern or a preference? An investigation into the ability of privacy scales to capture and distinguish granular privacy constructs},
  author={Colnago, Jessica and Cranor, Lorrie Faith and Acquisti, Alessandro and Stanton, Kate Hazel},
  booktitle={Eighteenth symposium on usable privacy and security (SOUPS 2022)},
  pages={331--346},
  year={2022}
}

@article{wu2025towards,
  title={Towards automating data access permissions in ai agents},
  author={Wu, Yuhao and Yang, Ke and Roesner, Franziska and Kohno, Tadayoshi and Zhang, Ning and Iqbal, Umar},
  journal={arXiv preprint arXiv:2511.17959},
  year={2025}
}

@article{zufferey2025ai,
  title={“AI is from the devil.” Behaviors and Concerns Toward Personal Data Sharing with LLM-based Conversational Agents},
  author={Zufferey, No{\'e} and Abdelwahab Gaballah, Sarah and Marky, Karola and Zimmermann, Verena},
  journal={Proceedings on Privacy Enhancing Technologies},
  volume={2025},
  number={3},
  pages={5--28},
  year={2025},
  publisher={Privacy Enhancing Technologies Symposium Advisory Board}
}

@inproceedings{liu2025prevalence,
  title={Prevalence Overshadows Concerns? Understanding Chinese Users' Privacy Awareness and Expectations Towards LLM-Based Healthcare Consultation},
  author={Liu, Zhihuang and Hu, Ling and Zhou, Tongqing and Tang, Yonghao and Cai, Zhiping},
  booktitle={2025 IEEE Symposium on Security and Privacy (SP)},
  pages={2716--2734},
  year={2025},
  organization={IEEE}
}

@article{chhikara2025mem0,
  title={Mem0: Building production-ready ai agents with scalable long-term memory},
  author={Chhikara, Prateek and Khant, Dev and Aryan, Saket and Singh, Taranjeet and Yadav, Deshraj},
  journal={arXiv preprint arXiv:2504.19413},
  year={2025}
}

@article{xu2025mem,
  title={A-mem: Agentic memory for llm agents},
  author={Xu, Wujiang and Liang, Zujie and Mei, Kai and Gao, Hang and Tan, Juntao and Zhang, Yongfeng},
  journal={arXiv preprint arXiv:2502.12110},
  year={2025}
}

@inproceedings{yang2023large,
  title={Large language models as optimizers},
  author={Yang, Chengrun and Wang, Xuezhi and Lu, Yifeng and Liu, Hanxiao and Le, Quoc V and Zhou, Denny and Chen, Xinyun},
  booktitle={The Twelfth International Conference on Learning Representations},
  year={2023}
}

\clearpage
\onecolumn


\appendix

\section{Prompts for Evaluation}
We provide the set of prompts that were used in our experimental evaluations below.

\subsection{Zero-shot Prompts}\label{app:zero_shot_prompts}
\begin{prompt}[fig:zero_shot_prompt]{Zero-shot prompt for SPA dataset}
Role: You are an AI agent expert in applying the theory of Contextual Integrity (CI) to make privacy judgments on information-sharing requests on behalf of a user.

Objective: Analyze an Incoming Request then judge if it is "appropriate" or "inappropriate" based on your knowledge of general privacy norms.

Information Flow Request
Each request is broken down into five fields from CI that describe a specific flow of information:

data_type: The piece of information being shared.

data_subject: The owner of the data.

data_sender: The individual/entity sharing the data.

data_recipient: The individual/entity receiving the data.

transmission_principle: The purpose and/or safeguard condition under which the data is shared.

For this task, assume the data_subject is always the 'user' and the data_sender is always the 'assistant provider'.

Required Output Format
Your entire output must be a single JSON object. Do not include any text outside of this JSON object. The structure must be as follows:

{{
  "judgment": "appropriate or inappropriate"
}}

Incoming Request to Judge:
{incoming_request}

Output:

\end{prompt}
\begin{prompt}[fig:education_zero_shot_prompt]{Zero-shot prompt for Education dataset}
Role: You are an AI agent expert in applying the theory of Contextual Integrity (CI) to make privacy judgments on information-sharing requests on behalf of a user.

Objective: Analyze an Incoming Request then judge if it is "appropriate" or "inappropriate" based on your knowledge of general privacy norms.

Information Flow Request
Each user request is broken down into five fields that describe a specific flow of information:

data_type: The piece of information being shared.

data_subject: The owner of the data.

data_sender: The individual/entity sharing the data.

data_recipient: The individual/entity receiving the data.

transmission_principle: The condition under which the data is shared.

For this task, assume the data_subject and the user is always the 'student'.

Required Output Format
Your entire output must be a single JSON object. Do not include any text outside of this JSON object. The structure must be as follows:

{{
  "judgment": "appropriate or inappropriate"
}}

Incoming Request to Judge:
{incoming_request}

Output:
    
\end{prompt}

\subsection{Privacy Norms Prompts}\label{app:priv_norms_prompts}
\begin{prompt}[fig:spa_general_norms]{Privacy norms prompt for SPA dataset}
Role: You are an AI agent expert in applying the theory of Contextual Integrity (CI) to make privacy judgments on information-sharing requests on behalf of a user.

Objective: Analyze an Incoming Request then judge if it is "appropriate" or "inappropriate" based on a provided set of general privacy norms.

Information Flow Request
Each request is broken down into five fields from CI that describe a specific flow of information:

data_type: The piece of information being shared.

data_subject: The owner of the data.

data_sender: The individual/entity sharing the data.

data_recipient: The individual/entity receiving the data.

transmission_principle: The purpose and/or safeguard condition under which the data is shared.

For this task, assume the data_subject is always the 'user' and the data_sender is always the 'assistant provider'.

General Privacy Norms:
Define user recipients to be a data_recipient from the following set:

parents, partner, siblings, children, housemates, close friends, close family, neighbours, house helper/keeper, visitors in general.

Additionally, define non-user recipients to be a data_recipient from the following set:

assistant provider, third party skills/actions, advertising agencies, law enforcement agencies

Then we provide the top appropriate and inappropriate privacy norms below. Each rule contains a subset of the CI fields with its corresponding value.
If a norm does not contain a CI field (e.g., transmission_principle), then that field can take on any value (e.g., transmission_principle can be any value).

The following are privacy norms for user recipients:

N1. data_type: 'sleeping hours', judgment: 'inappropriate'
N2. data_type: 'contacts', judgment: 'inappropriate'
N3. data_type: 'voice command history', judgment: 'inappropriate'
N4. data_type: 'email content', judgment: 'inappropriate'
N5. data_type: 'bank account details', judgment: 'inappropriate'
N6. data_recipient: 'neighbours', judgment: 'inappropriate'
N7. data_recipient: 'visitors in general', judgment: 'inappropriate'

The following are privacy norms for non-user recipients:

N1 data_type: 'contacts', judgment: 'inappropriate'
N2 data_type: 'video calls', judgment: 'inappropriate'
N3 data_type: 'bank account details', judgment: 'inappropriate'
N4 transmission_principle: 'no purpose/condition', judgment: 'inappropriate'
N5 data_recipient: 'non-relevant skills/actions', judgment: 'inappropriate'
N6 data_recipient: 'advertising agencies', judgment: 'inappropriate'

Required Output Format
Your entire output must be a single JSON object. Do not include any text outside of this JSON object. The structure must be as follows:

{{
  "judgment": "appropriate or inappropriate"
}}

Incoming Request to Judge:
{incoming_request}

Output:

\end{prompt}

\begin{prompt}[fig:education_general_norms_prompts]{Privacy norms prompt for Education Dataset}
Role: You are an AI agent expert in applying the theory of Contextual Integrity (CI) to make privacy judgments on information-sharing requests on behalf of a user.

Objective: Analyze an Incoming Request then judge if it is "appropriate" or "inappropriate" based on a provided set of general privacy norms.

Information Flow Request
Each user request is broken down into five fields that describe a specific flow of information:

data_type: The piece of information being shared.

data_subject: The owner of the data.

data_sender: The individual/entity sharing the data.

data_recipient: The individual/entity receiving the data.

transmission_principle: The condition under which the data is shared.

For this task, assume the data_subject and the user is always the 'student'.

General Privacy Norms:
We provide the top appropriate and inappropriate privacy norms below. Each privacy norm contains a subset of the CI fields with its corresponding value

The top six appropriate privacy norms:
N1. data_sender: 'professor', data_recipient: 'graduate schools', data_type: 'attendance', transmission_principle: 'with subject's consent'
N2. data_sender: 'registrar', data_recipient: 'parents', data_type: 'grades', transmission_principle: 'with subject's consent'
N3. data_sender: 'professor', data_recipient: 'department chair', data_type: 'term papers', transmission_principle: 'with subject's consent'
N4. data_sender: 'professor', data_recipient: 'registrar', data_type: 'participation', transmission_principle: 'with subject's consent'
N5. data_sender: 'librarian', data_recipient: 'department chair', data_type: 'photo', transmission_principle: 'with subject's consent'
N6. data_sender: 'registrar', data_recipient: 'graduate schools', data_type: 'grades', transmission_principle: 'with subject's consent'

The top five inappropriate privacy norms:
N1. data_sender: 'TA', data_recipient: 'classmates', data_type: 'grades', transmission_principle: 'if the subject is performing poorly'
N2. data_sender: 'TA', data_recipient: 'classmates', data_type: 'transcript', transmission_principle: 'with the requirement of confidentiality'
N3. data_sender: 'TA', data_recipient: 'librarian', data_type: 'grades', transmission_principle: 'with a request from the subject'
N4. data_sender: 'advisor', data_recipient: 'classmates', data_type: 'transcript', transmission_principle: 'with the requirement of confidentiality'
N5. data_sender: 'registrar', data_recipient: 'classmates', data_type: 'transcript', transmission_principle: 'with a request from the subject'

Required Output Format
Your entire output must be a single JSON object. Do not include any text outside of this JSON object. The structure must be as follows:

{{
  "judgment": "appropriate or inappropriate"
}}

Incoming Request to Judge:
{incoming_request}

Output:
 
\end{prompt}

\clearpage

\subsection{ICL Prompts}\label{app:icl_prompts}
\begin{prompt}[fig:spa_icl]{ICL prompt SPA dataset}
Role: You are an AI agent expert in applying the theory of Contextual Integrity (CI) to make privacy judgments on information-sharing requests on behalf of a user.

Ojective: Analyze an Incoming Request and judge if it is "appropriate" or "inappropriate" based on the user's prior privacy judgments.

Information Flow Request
Each request is broken down into five fields that describe a specific flow of information:

data_type: The piece of information being shared.

data_subject: The owner of the data.

data_sender: The individual/entity sharing the data.

data_recipient: The individual/entity receiving the data.

transmission_principle: The purpose and/or safeguard condition under which the data is shared.

For this task, assume the data_subject is always the 'user' and the data_sender is always the 'assistant provider'.

History of Prior Requests with User Judgments (this list can be empty or contain multiple entries):
{icl_examples}

Step-by-Step Instructions:
1. Analyze the history of prior requests with user judgments to understand the user's privacy preferences.
Look for patterns across all five fields. For example, note which data_recipients are trusted with a certain data_type,
or under which transmission_principle sensitive information is allowed to be shared.
2. Compare the Incoming Request to these established user preferences to determine the judgment of the incoming request.

Required Output Format
Your entire output must be a single JSON object. Do not include any text outside of this JSON object.
The structure must be as follows:

{{
  "judgment": "appropriate or inappropriate"
}}

Incoming Request to Judge:
{incoming_request}

Output:
\end{prompt}
\begin{prompt}[fig:education_icl_prompt]{ICL prompt for Education dataset}
Role: You are an AI agent expert in applying the theory of Contextual Integrity (CI) to make privacy judgments on information-sharing requests on behalf of a user.

Ojective: Analyze an Incoming Request and judge if it is "appropriate" or "inappropriate" based on the user's prior privacy judgments.

Information Flow Request
Each user request is broken down into five fields that describe a specific flow of information:

data_type: The piece of information being shared.

data_subject: The owner of the data.

data_sender: The individual/entity sharing the data.

data_recipient: The individual/entity receiving the data.

transmission_principle: The condition under which the data is shared.

For this task, assume the data_subject and the user is always the 'student'.

History of Prior Requests with User Judgments (this list can be empty or contain multiple entries):
{icl_examples}

Step-by-Step Instructions:
1. Analyze the history of prior requests with user judgments to understand the user's privacy preferences.
Look for patterns across all five fields. For example, note which data_recipients are trusted with a certain data_type,
or under which transmission_principle sensitive information is allowed to be shared.
2. Compare the Incoming Request to these established user preferences to determine the judgment of the incoming request.

Required Output Format
Your entire output must be a single JSON object. Do not include any text outside of this JSON object.
The structure must be as follows:

{{
  "judgment": "appropriate or inappropriate"
}}

Incoming Request to Judge:
{incoming_request}

Output:

\end{prompt}

\subsection{ICL (w Undet) Propmts}\label{app:icl_w_undet_prompts}
\begin{prompt}[fig:spa_icl_undetermined]{ICL (w Undetermined) prompt for SPA dataset}
Role: You are an AI agent expert in applying the theory of Contextual Integrity (CI) to make privacy judgments on information-sharing requests on behalf of a user.

Ojective: Analyze an Incoming Request and judge if it is "appropriate", "inappropriate", or "undetermined" based on the user's prior privacy judgments.

Information Flow Request
Each request is broken down into five fields that describe a specific flow of information:

data_type: The piece of information being shared.

data_subject: The owner of the data.

data_sender: The individual/entity sharing the data.

data_recipient: The individual/entity receiving the data.

transmission_principle: The purpose and/or safeguard condition under which the data is shared.

For this task, assume the data_subject is always the 'user' and the data_sender is always the 'assistant provider'.

History of Prior Requests with User Judgments (this list can be empty or contain multiple entries):
{icl_examples}

Step-by-Step Instructions:
1. Analyze the history of prior requests with user judgments to understand the user's privacy preferences.
Look for patterns across all five fields. For example, note which data_recipients are trusted with a certain data_type,
or under which transmission_principle sensitive information is allowed to be shared.
2. Compare the Incoming Request to these established user preferences to determine the judgment of the incoming request.
If you are unable to confidently judge whether the Incoming Request is "appropriate" or "inappropriate" based on the Prior Requests,
then your judgment should be "undetermined".

Required Output Format
Your entire output must be a single JSON object. Do not include any text outside of this JSON object.
The structure must be as follows:

{{
  "judgment": "appropriate, inappropriate, or undetermined"
}}

Incoming Request to Judge:
{incoming_request}


Incoming Request to Judge:
{incoming_request}

Output:
    
\end{prompt}
\begin{prompt}[fig:education_icl_undetermined_prompt]{ICL (w Undetermined) prompt for Education dataset}
Role: You are an AI agent expert in applying the theory of Contextual Integrity (CI) to make privacy judgments on information-sharing requests on behalf of a user.

Ojective: Analyze an Incoming Request and judge if it is "appropriate", "inappropriate", or "undetermined" based on the user's prior privacy judgments.

Information Flow Request
Each user request is broken down into five fields that describe a specific flow of information:

data_type: The piece of information being shared.

data_subject: The owner of the data.

data_sender: The individual/entity sharing the data.

data_recipient: The individual/entity receiving the data.

transmission_principle: The condition under which the data is shared.

For this task, assume the data_subject and the user is always the 'student'.

History of Prior Requests with User Judgments (this list can be empty or contain multiple entries):
{icl_examples}

Step-by-Step Instructions:
1. Analyze the history of prior requests with user judgments to understand the user's privacy preferences.
Look for patterns across all five fields. For example, note which data_recipients are trusted with a certain data_type,
or under which transmission_principle sensitive information is allowed to be shared.
2. Compare the Incoming Request to these established user preferences to determine the judgment of the incoming request.
If you are unable to confidently judge whether the Incoming Request is "appropriate" or "inappropriate" based on the Prior Requests,
then your judgment should be "undetermined".

Required Output Format
Your entire output must be a single JSON object. Do not include any text outside of this JSON object.
The structure must be as follows:

{{
  "judgment": "appropriate, inappropriate, or undetermined"
}}

Incoming Request to Judge:
{incoming_request}


Incoming Request to Judge:
{incoming_request}

Output:

\end{prompt}

\clearpage

\subsection{\method{} Prompts}\label{app:method_prompts}
\begin{prompt}[fig:ontology_gen]{Ontology generation prompt for SPA dataset}
Role:
You are an AI expert specializing in data privacy, contextual integrity (CI), and ontological modeling.

Task:
Your task is to generate three hierarchical ontologies: one for data_type, one for data_recipient, and one for transmission_principle. These ontologies must be usable by a language model to logically map all possible values for each field into ranked levels.

The hierarchy for each ontology must be based on two sources, in order of priority:

1. User Privacy Preferences: The specific judgments in the Information flows are the primary driver for creating levels. Any two values (e.g., value_A vs. value_B) that are treated differently by the user in the judgments must be placed in separate, distinct levels.

2. General Norms: Use general norms only after all user judgments are satisfied. General norms should be used to (a) write a clear description for the levels you've already created and (b) logically sort other common values related to the Smart House Personal Assistant (SPA) ecosystem into the levels defined by the user's preferences.

Core Concepts: Contextual Integrity (CI) A flow of information is described by five CI fields:

data_type: The piece of information being shared.

data_subject: The owner of the data.

data_sender: The individual/entity sharing the data.

data_recipient: The individual/entity receiving the data.

transmission_principle: The purpose and/or safeguard condition under which the data is shared. The format is: purpose, condition. Note that if transmission_principle is empty, it should be interpreted as having no purpose.

For this task, assume the data_subject is always the 'user' and the data_sender is always the 'assistant provider'. Your ontologies will define the values for the other three fields.

Input: User Privacy Judgments.

You will be provided with a list of the user's judgments on specific information flows. These flows determine the user's specific preferences.

Each scenario is structured as: (data_type, data_recipient, transmission_principle) -> judgment (where judgment is 'appropriate' or 'inappropriate')

Information flows:
{privacy_preferences}

Ontology Generation Rules

You must generate one ontology for each of the three fields below.

1. data_type Ontology
Relationship: The Data Sensitivity as percieved by the user.

Hierarchy: Level 1 (L1) must contain the least sensitive data types. Sensitivity increases at higher levels .

Value Sources: Base your categories on common SPA data types. Use values from data_type in Information flows, then expand this by referencing common categories from Google Actions and Alexa Skills.

2. data_recipient Ontology
Relationship: The degree of Trust by the user to the recipient. Additionally, the authority each recipient has to receive the data.

Hierarchy: Level 1 must contain the most trusted recipients and/or recipients with most authority, with trust and/or authority decreasing at higher levels.

Value Sources: Use the values from data_recipient in Information flows, then expand this to include a comprehensive set of potential recipients in an SPA ecosystem, such as household members or visitors. Additionally, entities who may have access to the user's SPA ecosystem, such as providers.

3. transmission_principle Ontology
Relationship: The combination of (A) The Purpose of the transmission and (B) The Condition (Safeguard) applied.

Hierarchy: The hierarchy must rank combinations of these two factors. L1 represents the most appropriate combination (e.g., a related purpose with strong safeguards). Purpose and safeguard decrease at higher levels (e.g., no purpose and/or weaker/no safeguard).

Key Distinction: The hierarchy must separate principles based on the strength of the safeguard (Condition). A principle with a specific safeguard (e.g., 'in an anonymized form', 'with explicit user consent') must be in a higher (more appropriate, lower L-number) level than the exact same purpose with a weaker or no safeguard.

Value Sources: Use the principles in Information flows as your starting point. Expand this to include a comprehensive set of combinations involving purpose and safety of the data.

Coverage and Exclusivity Requirement:

1. Mutually Exclusive (The "No Overlap" Rule): This is your most important rule. The description for each level must be precise enough that any given value maps to exactly one level. Levels must not overlap.

2. Collectively Exhaustive (The "Full Coverage" Rule): Your generated ontologies must be comprehensive. They must logically cover all values present in the Information flows input, as well as all other common values for each CI field in an SPA context.

3. Judgment-Driven Grouping: To achieve both rules, use the Information flows to define your levels.
 - Example, consider two scenarios that differ by data_recipient: (..., recipient_A, ...) is appropriate but (..., recipient_B, ...) is inappropriate. Therefore, 'recipient_A' and 'recipient_"B' must be in different levels. Your level description must capture why (e.g., "L1: Description of trusted group, e.g., recipient_A" vs. "L4: Description of untrusted group, e.g., recipient_B").

Output Format:
Your output must consist only of the three generated ontologies. Do not add any preamble, explanation, or conversational text. Use the following format:

data_type Ontology
L1. "Description of this level"
L2. "Description of this level"
...

data_recipient Ontology
L1. "Description of this level"
L2. "Description of this level"
...

transmission_principle Ontology
L1. "Description of this level"
L2. "Description of this level"
...

Output:

\end{prompt}
\begin{prompt}[fig:education_ontology_gen]{Ontology generation prompt for Education dataset}
Role:
You are an AI expert specializing in data privacy, contextual integrity (CI), and ontological modeling.

Task:
Your task is to generate four hierarchical ontologies: one for data_type, one for data_sender, one for data_recipient, and one for transmission_principle. These ontologies must be usable by a language model to logically map all possible values for each field into ranked levels.

The hierarchy for each ontology must be based on two sources, in order of priority:

1. User Privacy Preferences: The specific judgments in the Information flows are the primary driver for creating levels. Any two values (e.g., value_A vs. value_B) that are treated differently by the user in the judgments must be placed in separate, distinct levels.

2. General Norms: General privacy norms and common privacy expecations around school/education setting. Use general norms only after all user judgments are satisfied. They should be used to (a) write a clear description for the levels you've already created and (b) logically sort other common values related to the school/education context into the levels defined by the user's preferences.

Core Concepts: Contextual Integrity (CI) A flow of information is described by five CI fields:

data_type: The piece of information being shared.

data_subject: The owner of the data.

data_sender: The individual/entity sharing the data.

data_recipient: The individual/entity receiving the data.

transmission_principle: The condition under which the data is shared.

For this task, assume the data_subject and the user is always the 'student'. Your ontologies will define the values for the other four fields.

Input: User Privacy Judgments.

You will be provided with a list of the user's judgments on specific information flows. These flows determine the user's specific preferences.

Each scenario is structured as: (data_type, data_sender, data_recipient, transmission_principle) -> judgment (where judgment is 'appropriate' or 'inappropriate')

Information flows:
{privacy_preferences}

Ontology Generation Rules

You must generate one ontology for each of the four fields below.

1. data_type Ontology

Relationship: The student's perceived Data Sensitivity of data_type.

Hierarchy: Level 1 must contain the least sensitive data types, with sensitivity increasing at higher levels.

Value Sources: Use values from data_type in Information flows, then expand this by referencing common school/education data types.

2. data_sender Ontology

Relationship: The authority of the data_sender to send the student's data.

Hierarchy: Level 1 must contain the senders with the most authority, with authority decreasing at higher levels.

Value Sources: Use the values from data_sender in Information flows, then expand this to include a comprehensive set of potential recipients in a school/education setting, such as professors.

3. data_recipient Ontology

Relationship: The degree of Trust a student typically has in these people. Additionally, the authority each recipient has to receive the data.

Hierarchy: Level 1 must contain the most trusted recipients and/or recipients with most authority, with trust and/or authority decreasing at higher levels.

Value Sources: Use the values from data_recipient in Information flows, then expand this to include a comprehensive set of potential recipients in a school/education setting, such as classmates.

4. transmission_principle Ontology

Relationship: The degree of Data Safety offered to the student.

Hierarchy: Level 1 must represent the highest degree of data safety, with the degree of safety decreasing at higher levels.

Value Sources: Use the principles in Information flows as your starting point. Expand this to include other conditions involved in a school/education setting.

Coverage and Exclusivity Requirement:

1. Mutually Exclusive (The "No Overlap" Rule): This is your most important rule. The description for each level must be precise enough that any given value maps to exactly one level. Levels must not overlap.

2. Collectively Exhaustive (The "Full Coverage" Rule): Your generated ontologies must be comprehensive. They must logically cover all values present in the Information flows input, as well as all other common values for each CI field in a school/education context.

3. Judgment-Driven Grouping: To achieve both rules, use the Information flows to define your levels.
 - Example, consider two scenarios that differ by data_recipient: (..., recipient_A, ...) is appropriate but (..., recipient_B, ...) is inappropriate. Therefore, 'recipient_A' and 'recipient_B' must be at different levels. Your level description must capture why (e.g., "L1: Description of trusted group, e.g., recipient_A" vs. "L4: Description of untrusted group, e.g., recipient_B").

Output Format:
Your output must consist only of the three generated ontologies. Do not add any preamble, explanation, or conversational text. Use the following format:

data_type Ontology
L1. "Description of this level"
L2. "Description of this level"
...

data_recipient Ontology
L1. "Description of this level"
L2. "Description of this level"
...

transmission_principle Ontology
L1. "Description of this level"
L2. "Description of this level"
...

Output:

\end{prompt}
\begin{prompt}[fig:ontology_map_prompt]{Ontology mapping prompt}
Role:
You are an AI expert specializing in data privacy, contextual integrity (CI), and ontological modeling.

Task:
Your task will be to map string values to the relevant level of an ontology.

Background:
A request is an information flow that can be described by five CI fields:

data_type: The piece of information being shared.

data_subject: The owner of the data.

data_sender: The entity sharing the data.

data_recipient: The entity receiving the data.

transmission_principle: The purpose and/or safeguard condition under which the data is shared. The format is: purpose, condition. Note that if transmission_principle is empty, it should be interpreted as having no purpose.

You will be provided a prior request, an incoming request, a list of ontologies, and the CI field that differs between the prior and incoming request.

Step-by-step instructions:
1. Identify the field name from the Differing Field input.

2. Get the value for this field from Prior Request. This will be the prior_A value.

3. Get the value for this field from Incoming Request. This will be the incoming_B value.

4. Select the ontology from Ontologies that matches the differing field.

5. Carefully analyze the descriptions of each level (L1, L2, etc.) in the selected ontology.

6. Map the prior_A value to its most accurate level (e.g., "L1"). This will be the mapped_prior_A value.

7. Map the incoming_B value to its most accurate level (e.g., "L2"). This will be the mapped_incoming_B value.

Mapping Requirements:
Analyze each level's description carefully to find the best fit.

You must map both values to a level. If a value is ambiguous or doesn't seem to fit perfectly, select the best-fit level. Do not refuse to map.

Output Format:
Your output must be a single, raw JSON object. Nothing else.

Do not include any other text, explanations, preamble, or markdown formatting (like ```json). The JSON object must be the only thing you output.

The JSON object must use these exact keys and format:

{{
  "prior_A": "differing field value from prior request obtained in Step 1.",
  "incoming_B": "differing field value from incoming request obtained in Step 1.",
  "mapped_prior_A": "prior_A mapped to the level of the selected ontology in Step 3.",
  "mapped_incoming_B": "incoming_B mapped to the level of the selected ontology in Step 3."
}}

**EXAMPLES:

Correct output (Good):

{{
  "prior_A": "credit card information",
  "incoming_B": "birth date",
  "mapped_prior_A": "L3",
  "mapped_incoming_B": "L1"
}}

**Incorrect Example (Bad - DO NOT DO THIS):**
{{
  "prior_A": "some value",
  - "incoming_B": "another value",  <-- DO NOT ADD A HYPHEN
  "mapped_prior_A": "L1",
  "mapped_incoming_B": "L2"
}}

Ontologies:
{ontologies}

Prior Request:
{prior_request}

Incoming Request:
{incoming_request}

Differing Field:
{differing_field}

**CRITICAL REMINDER: Your response must be *only* the syntactically valid JSON object as specified in the "Output Format". It must start with `{{` and end with `}}`. Do not include any other text, markdown, or **hyphens (`-`).**

Output:

\end{prompt}

\end{document}